\documentclass[aps,prl,longbibliography,twocolumn,superscriptaddress,showpacs]{revtex4-2}

\usepackage{braket}
\usepackage{graphicx}
\usepackage{bm}
\usepackage{bbm}
\usepackage{color}
\usepackage[page,header]{appendix}

\usepackage{booktabs,tabularx}
\usepackage{dcolumn} 
\usepackage{comment}
\usepackage{epstopdf}
\usepackage{bbold}

\usepackage[colorlinks=true,linkcolor=blue,urlcolor=blue,citecolor=blue]{hyperref}
\usepackage{commath}
\usepackage[caption=false]{subfig}
\usepackage{afterpage}

\usepackage{soul}

\definecolor{orange}{RGB}{255,127,200}

\DeclareFontFamily{U}{mathx}{\hyphenchar\font45}
\DeclareFontShape{U}{mathx}{m}{n}{
      <5> <6> <7> <8> <9> <10>
      <10.95> <12> <14.4> <17.28> <20.74> <24.88>
      mathx10
      }{}
\DeclareSymbolFont{mathx}{U}{mathx}{m}{n}
\DeclareFontSubstitution{U}{mathx}{m}{n}
\DeclareMathAccent{\widecheck}{0}{mathx}{"71}
\DeclareMathAccent{\wideparen}{0}{mathx}{"75}

\begin{document}

\title{A theory of Stimulated and Spontaneous Axion Scattering}
\author{M. Smith}
\affiliation{Materials Science Division, Argonne National Laboratory, Lemont, Illinois 60439, USA}
\author{Kartiek Agarwal}
\affiliation{Materials Science Division, Argonne National Laboratory, Lemont, Illinois 60439, USA}
\author{Ivar Martin}
\affiliation{Materials Science Division, Argonne National Laboratory, Lemont, Illinois 60439, USA}
\date{\today}

\begin{abstract}
We present a theory for nonlinear, resonant excitation of dynamical axions by counter-propagating electromagnetic waves in materials that break both $\mathcal{P}$ and $\mathcal{T}$ symmetries. We show that dynamical axions can mediate an exponential growth in the amplitude of the lower frequency (Stokes) beam. We also discuss spontaneous generation of a counter-propagating Stokes mode, enabled by resonant amplification of quantum and thermal fluctuations in the presence of a single pump laser. Remarkably, the amplification can be orders of magnitude larger than that obtained via stimulated Brillouin and Raman scattering processes, and can be modulated with the application of external magnetic fields, making stimulated axion scattering promising for optoelectronics applications.
\end{abstract}

\maketitle

\addtolength{\abovedisplayskip}{-1mm}

\paragraph{Introduction.---}The propagation of electromagnetic (EM) fields through materials is governed by the celebrated Maxwell's equations. In vacuum, these equations arise from the usual electromagnetic action $\mathcal S_{EM} = \frac{1}{2}\int d^3x dt \left(\epsilon_0 \bm{E}^2 - \bm{B}^2/\mu_0\right)$, where $\bm{E}$ and $\bm{B}$ are the electric and magnetic fields and $\epsilon_0$ and $\mu_0$ are the vacuum permittivity and permeability, respectively. For waves in a medium, it often suffices to promote these constants to material dependent values without changing the form of the action. However, in materials that break  time reversal ($\mathcal{T}$) and  inversion ($\mathcal{P}$) symmetries,  a novel term becomes possible
\begin{align}\label{eq:Sa}
    \mathcal S_a = -\frac{g_{a\gamma\gamma}}{4}\int d^3x dt\ \theta\ \bm E \cdot \bm B, 
\end{align}
with a  pseudoscalar field $\theta$  (odd under 
both $\mathcal{T}$ and $\mathcal{P}$) coupled to electromangetic fields with strength $g_{a\gamma\gamma}$. 
This unusual term first appeared in quantum chromodynamics in the context of  the strong charge conjugation-parity problem ~\cite{peccei_cp_1977,weinberg_new_1978,wilczek_problem_1978}. 
The elementary excitations of the $\theta$ field, ``axions", are one of the prime candidates for cosmological dark matter ~\cite{wilczek_two_1987}.  Despite significant efforts~\cite{sikivie_experimental_1983, rybka_direct_2014,budker_proposal_2014,brubaker_first_2017,madmax_working_group_dielectric_2017,irastorza_new_2018,goryachev_axion_2018,lawson_tunable_2019,roising_axion-matter_2021}, they are yet to be detected experimentally. 

The magnetoelectric coupling (\ref{eq:Sa}) has been known to exist in materials like Cr$_2$O$_3$  for quite some time~\cite{wiegelmann_magnetoelectric_1994,coh_chern-simons_2011,wu_quantized_2016}. The corresponding action, $\mathcal S_{\text{ME}} = \int d^3 x dt\ \tilde\alpha_{ij} E_i B_j$, typically requires breaking both $\mathcal{P}$ and $\mathcal{T}$ symmetries. However, recently it has been recognized that even in the presence of $\mathcal P$ and/or $\mathcal T$, such a coupling can exist, if $\tilde\alpha_{ij}$ is given by $ \alpha \epsilon_0 c \theta \delta_{ij} /\pi $, with $c$  the speed of light, $\alpha$ the fine structure constant, and $\theta$ is an integer multiple of $\pi$. Having  $\theta = n \pi$ with $n$ odd is the defining characteristic of topological insulators and is the origin of the topological magnetoelectric effect~\cite{qi_topological_2008,zhang_topological_2009,li_dynamical_2010,tse_giant_2010}.  


When both $\mathcal P$ and $\mathcal T$  symmetries are broken, $\theta$ is no longer quantized. This is the case in Cr$_2$O$_3$, but also some magnetically doped topological insulators ~\cite{li_dynamical_2010,sekine_axion_2021}. In the latter, finite antiferromagnetic (Neel) order leads to the deviation of $\theta$ from $\pi$, and longitudinal fluctuations of the Neel order parameter translate into fluctuations of $\theta$. 
To avoid confusion with the static part of $\theta$, these fluctuations are sometimes referred to as {\em dynamical axions} (DAs). DAs can hybridize with photons, forming axion-polaritons, and can manifest  in a variety of dynamical responses 
~\cite{sekine_chiral_2016,taguchi_electromagnetic_2018,liu_anisotropic_2020, ahn_theory_2022, liebman_multiphoton_2023,gao_detecting_2024}. Recent experimental work in thin film MnBi$_2$Te$_4$, which is predicted to host DAs, not only observed an axion-like coupling between the Neel order and $\bm E \cdot \bm B$
 ~\cite{gao_layer_2021} but also directly probed DAs~\cite{Qiu_Observation_2024}.

In this Letter, we focus on an inherently nonlinear phenomenon associated with  light propagation in media supporting DAs, namely  stimulated axion scattering (SAS).
We show that the nonlinear excitation of DAs by two counterpropagating and orthogonally polarized fields leads to energy transfer between the two waves, with the intensity of the lower frequency (Stokes) mode growing exponentially in space. This can lead to reflection and transmission coefficients of the Stokes mode to be greater than $1$, and provides a sensitive probe of axion dynamics in the material.  We also show that the Stokes mode can be generated spontaneously in the presence of a single applied electromagnetic wave when allowing for quantum or thermal fluctuations of the  Neel order. Apart from providing a method for axion spectroscopy, this latter phenomenon can be employed to realize the phase conjugation effect ~\cite{zeldovich_principles_1985}, which  can be used in holography and image correction~\cite{hillman_digital_2013}.

The exponential growth of the Stokes mode in SAS is similar to that seen in Stimulated Brillouin Scattering (SBS)~\cite{brillouin_diffusion_1922} and Stimulated Raman Scattering (SRS)~\cite{hellwarth_theory_1963}.  
Instead of axions, the oscillator modes that enable SBS and SRS are atomic or molecular vibrations. 
Due to their high sensitivity, SBS and SRS have found applications in microscopy and spectroscopy both in inroganic materials and biological systems ~\cite{koski_brillouin_2005,scarcelli_confocal_2008,ballmann_stimulated_2015,cheng_emerging_2021,ranjan_stimulated_2024}. Furthermore, the orthogonality of polarizations which maximizes SAS, concomitantly minimizes SBS and SRS; this sensitivity to the polarization can thus help isolate these different phenomena.  
%
Remarkably, even though the DA is driven by the product of the electric and magnetic fields, which one naively would expect to have weaker coupling to matter than the product of electric fields, the exponential amplification rate for SAS can be orders of magnitude larger than for SBS and SRS. 
This difference arises from the relatively low stiffness of the DA as compared to phonons, making DAs easier to excite.
The  strength  of SAS  and its tunability with external fields  point towards possible practical importance of materials that host DAs for various optoelectronics applications where SBS and SRS are currently employed  \cite{cerny_solid_2004,bai_stimulated_2018}. 

\begin{figure}[h!]   \includegraphics[width=0.9\linewidth]{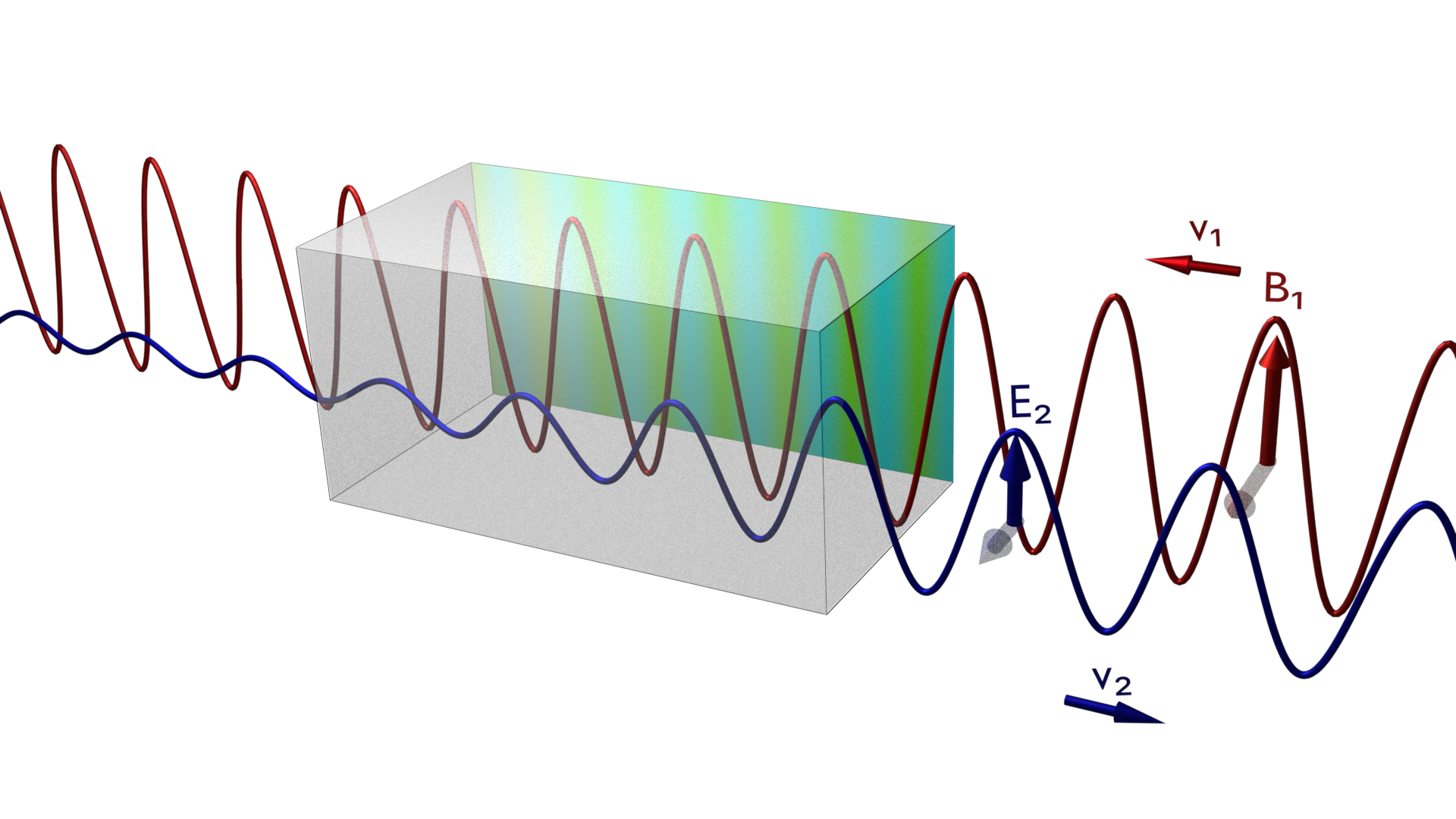}
    \caption{Counter-propagating and orthogonally polarized electromagnetic waves impinging on a sample hosting DAs. The lower frequency Stokes mode (blue) is amplified with the aid of the pump pulse (red) and excitation of DA field in the medium with the appropriate wavevector (green and blue). 
    }
    \label{fig:NLExcitation}
\end{figure}
{\em Equations of Motion.---}
The minimal action for the coupled EM fields and DAs in a material is given by
\begin{align}
    \mathcal S &= \int d^3 x dt \frac{1}{2}(\epsilon \bm{E}^2 - \bm{B}^2/\mu) + \tilde\alpha  \int d^3 x dt \theta \bm E \cdot \bm B \nonumber\\
    &+J  \int d^3 x dt \left[ (\partial_t \theta)^2 - (v \nabla \theta)^2 - m^2 \theta^2 \right].
\end{align}
Here we allow the permeability $\mu$ and permittivity $\epsilon$ to deviate from their vacuum values. $\tilde \alpha = \alpha \epsilon_0 c/\pi$ is the coupling between the EM fields and the DA, $J,\, v$ and $m$ are the stiffness, velocity, and mass of the DA, respectively. For the remainder of this letter, we neglect the static part of $\theta$, as it only weakly affects SAS (see discussion in SI).

Minimizing the action leads to the coupled equations of motion
\begin{subequations}
\begin{align}
\label{eq:EOME}
    \partial_t^2 \bm E - c'^2 \nabla^2 \bm E &= - \frac{1}{\epsilon} \partial_t \bm J_\theta, \\
\label{eq:EOMtheta}
    \partial_t^2 \theta + \gamma \partial_t \theta + v^2 \nabla^2 \theta + m^2 \theta &= \frac{\tilde\alpha}{2J}\bm E \cdot \bm B.    
\end{align}
\end{subequations}
Here $c' = 1/\sqrt{\mu \epsilon}$ is the speed of light in the material and the current due to DAs is given by $\bm J_\theta = \partial_t \bm P_\theta + \bm\nabla\times \bm M_\theta$,  where the polarization and magnetization induced by the DA are $\bm P_\theta =\tilde \alpha \theta \bm B$ and $\bm M_\theta = \tilde\alpha \theta \bm E$, respectively. A rate $\gamma$ has been introduced phenomenologically to describe the damping of axions (which in turn relates to the relaxation of appropriate spin waves in the material). 

As depicted in Fig.~\ref{fig:NLExcitation}, we consider two counter propagating electromagnetic waves, $ \bm E = \bm E_{1}(z,t) e^{i(\omega_1 t - k_1 z)} + \bm E_{2}(z,t) e^{i(\omega_2 t + k_2 z)}$ + c.c.,
where $\bm E_{1(2)}$ are the  slowly varying envelope functions and $
\omega_{1(2)}$ and $k_{1(2)}$ their frequency and wavevector in the DA medium, respectively; the waves are oriented such that $\bm E_{1(2)} \parallel \bm B_{2(1)}$ and propagating along the $z$-axis.  In the following we will ignore the spatial derivative part on the LHS of (\ref{eq:EOMtheta}), using the fact that the axion velocity is much smaller than the speed of light. Then, if the  difference of the incoming light beam frequencies $\Omega = \omega_1 - \omega_2$ matches  the DA energy $m$, DAs will be resonantly excited. We  refer to the higher frequency ($\omega_1$) mode as the pump, and the lower frequency ($\omega_2$) mode as the Stokes mode. We will also assume that $\Omega \ll \omega_{1,2}$, which allows us to use the rotating wave approximation. 
Making the ansatz $\theta = \theta_{12}(z,t) e^{i(\Omega t - Q z)} $ + c.c.,
with $Q = k_1 + k_2$ and 
plugging in the solution for $\theta_{12}$ from Eq.~\eqref{eq:EOMtheta} into Eq.~\eqref{eq:EOME}, we find for the envelope functions $E_{1,2}$ in the slowly varying envelope  approximation:
\begin{subequations}
\label{eqs:BulkE}
\begin{align}
    \nabla_z E_{1}+\frac{1}{c'} \partial_t E_{1} &= -\frac{i \omega_1}{c'^2}\frac{\tilde\alpha}{\epsilon}\theta_{12} E_{2}, \\
    \label{eq:BulkE2}
    \nabla_z E_{2} - \frac{1}{c'} \partial_t E_{2} &=  \frac{i\omega_2}{c'^2} \frac{\tilde\alpha}{\epsilon}\theta_{12}^* E_{1}.
\end{align}
\end{subequations}
In the limit when $E_{1}$ can be assumed constant ($|E_{1}|\gg|E_{2}|$), the RHS of Eq.~\eqref{eq:BulkE2} can be viewed as giving an  imaginary contribution to the index of refraction, which on resonance,  $\Omega = m$, goes as $-c' g_A  I_1/(2\omega_1) $, where $g_A$ is the gain coefficient  defined below in Eq.~\eqref{eq:AGain} and $I_1 = 2 \epsilon_0 c' |E_{1}|^2$ is the intensity of the pump.  This imaginary contribution to the index of refraction makes the amplitude of the Stokes mode $E_{2}$ grow exponentially as a function of $z$.

We can find time-independent solutions to the above equations for arbitrary incoming beam intensities and frequencies. Writing Eqs.~\eqref{eqs:BulkE} in terms of the intensites $I_i = 2 \epsilon_0 c' |E_{i}|^2$, their real parts yield 
\begin{subequations}
\label{eqs:I12}
\begin{align}
    \nabla_{z} I_1 &= - g_{A} I_1 I_2\\
    \nabla_z I_2 &= - g_{A} \xi_{12}I_1 I_2
\end{align}
\end{subequations}
where
\begin{align}
    g_A &= \frac{\alpha^2 n^4}{\pi^2 c^2} \cdot\frac{\epsilon_0}{\epsilon}\cdot \frac{\omega_1}{\gamma} \cdot \frac{1}{J \Omega} \cdot \frac{(\gamma \Omega)^2}{(m^2-\Omega^2)^2 + (\gamma \Omega)^2}.
\label{eq:AGain}
\end{align}
Here, $g_A$ is the gain coefficient for the SAS process mentioned above, $\xi_{12} = \omega_2/\omega_1$ and $n = \sqrt{\epsilon\mu/(\epsilon_0 \mu_0)}$ is the index of refraction of the DA medium. Note that Eqs.~(\ref{eqs:I12}) imply $\nabla_z I_1/\omega_1 = \nabla_z I_2/\omega_2$ which  can be interpreted as the conservation of photon number between the Stokes and pump fields (single absorbed pump photon converts into single Stokes photon and DA). 
The pump beam enters the sample at $z = 0$, while the Stokes beam at the opposite end at $z = L$. 
In the limit $\xi_{12} I_1(0) \gg I_2(0)$, the output Stokes intensity is readily obtained and found to grow exponentially in both the pump intensity and sample length:
\begin{align}
\label{eq:I2Bulk}
    I_2(0) &\approx I_2(L) e^{G_A}, \quad G_A \equiv g_A \xi_{12} I_1(0) L.
\end{align}
The pump intensity $I_1(L)$ can be obtained using the photon conservation equation; a general solution for arbitrary pump and Stokes intensities is given in the SI.

{\em Effect strength estimate. }
We now estimate $g_A$ using model parameters for Bi$_2$Se$_3$ doped with Fe~\cite{zhang_topological_2009}. The axion stiffness $J$ is given by (see SI for more details)
\begin{align}
\label{eq:J}
    J = M_0^2 \int\frac{d^3k}{(2\pi)^3}\frac{\sum_{i=1}^4 d_i^2}{16 \left( \sum_{i=1}^5 d_i^2\right)^{5/2}} \approx \frac{10^{3/2}}{16 (2\pi)^3} \frac{M_0^2}{A_1 A_2^2}
\end{align}
where $d_i$ are parameters in an effective four-band model for the system, $A_{1(2)}$ are velocities and $M_0$ is the gap at the $\Gamma$ point.
%
Using $M_0 = -.03$ eV, such that the band gap is trivial, $A_1 = 2.2 $~eV$\cdot$\AA ~, $A_2 = 4.1$ eV$\cdot$\AA~\cite{zhang_topological_2009}, an axion mass $m = 2$ meV, magnon lifetime $\tau = 0.1$ ns, pump frequency $\omega_1 = 1.481$ THz, index of refraction $n = 1$, and assuming we are on resonance, we find $g_A = 11.1 \text{ m/GW}$.

Importantly, $g_A$ is proportional to the magnon lifetime and may be larger as magnon lifetimes can exceed $10-100$ ns~\cite{bayrakci_lifetimes_2013}. Conversely, the gain bandwidth is protortional to the magnon relaxation rate; see Eq.~(\ref{eq:AGain}). Thus, an increase of the bandwidth can be traded for a decrease in gain amplitude and vice versa. We also note that $g_A$ is inversely proportional to the Neel temperature, as $m \sim k_B T_N$ which can be modulated, for instance, by applying a magnetic field perpendicular to the AFM ordering direction. The controllability of gain parameters using external fields makes media supporting DAs potentially more versatile for non-linear amplification than materials where SBS, SRS or inversion breaking $\chi^{(2)}$ non-linearities~\cite{manzoni_design_2016} are employed. 

The estimated gain $g_A$ is at least an order of magnitude larger than that obtained with SBS or SRS in materials typically used for nonlinear optics~\cite{boyd_nonlinear_2020}. These processes are similar to SAS with the distinction that SBS and SRS involve lattice and molecular vibrations, respectively, in lieu of axions. The large SAS effect is at first surprising given that SBS and SRS proceed via a putatively stronger coupling $\sim \delta \epsilon \bm{E} \cdot \bm{E}$, where $\delta \epsilon$ is a change in the dielectric constant due to lattice or molecular vibrations while the former corresponds to a weaker coupling $\sim \theta \bm{E} \cdot \bm{B}$. In the SI, we argue that the reason for the strength of SAS in comparison to SBS (which in turn is generally stronger than SRS) is because of the lower stiffness of axions compared to phonons, making them easier to excite. Another process used for parameteric amplification when inversion symmetry is absent are $\chi^{(2)}$ non-linearities. Here, the gain coefficient scales as $\sim \sqrt{I}$ and is generically weaker than that for SAS, SBS, and SRS, but the gain bandwidth is  much broader (for a quantiative comparison, see Ref.~\cite{manzoni_design_2016}). In short, SAS is a strong effect that could be used for parameteric amplification depending on the precise requirements of gain, bandwidth, and wavelength of interest. 

Finally, we note that the maximum amplification obtainable is not without bounds. 
The  DA amplitude is bounded due to the finite amplitude of the underlying Neel order. The deviation in the static $\theta$ from $\pi$ in AFM topological insulators is given by $\delta \theta = -m_5/M_0$~\cite{sekine_chiral_2016}, where $m_5 \sim k_B T_N$ is the contribution of the longitudinal AFM order to the electronic parameter $d_5$ introduced above. This limits the amplitude of DA to $\abs{\theta(z)} \le \abs{m_5/M_0}$. 
%
%
Since $\theta(z)$ is excited by the product of the pump and Stokes fields, this puts a bound on the geometric mean of the local intensities, $\sqrt{I_1(z) I_2(z)} < 2 \alpha \omega_1 /(\pi c g_A)$. While this does not bound the relative gain ($I_2(0)/I_2(L)$), it asserts a bound on the maximum output Stokes intensity.



\begin{figure}[t]
    \includegraphics[width=0.9\linewidth]{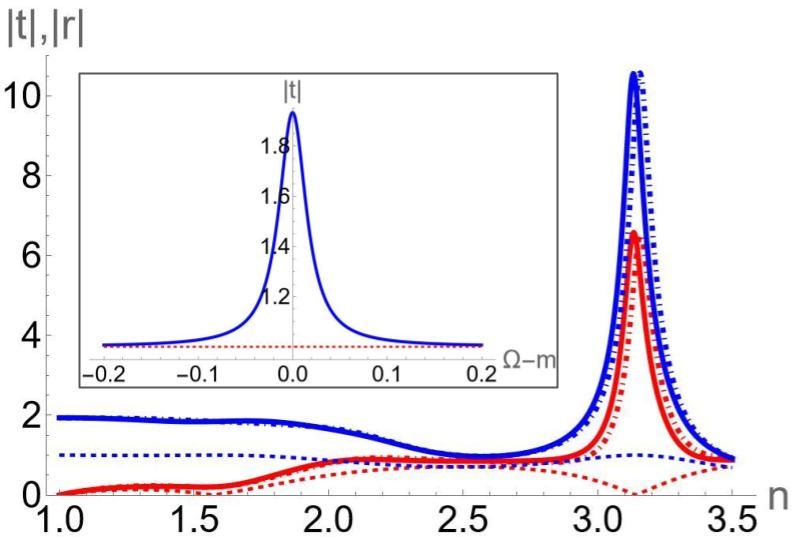}
    \caption{Reflectance $\abs{r}$ (red) and transmittance $\abs{t}$ (blue) for the Stokes mode plotted as a function of refractive index $n$, obtained numerically for a finite sample of length $L = 600$ $\mu$m with a gain factor of $G_A \approx 1.32$ at $n=1$. Solid (dashed) lines correspond to simulations in the presence (absence) of DAs. Dot-dashed lines are obtained by geometrically summing amplitudes of all possible light paths accounting for gain and phase accrued in propagating inside the sample. Inset shows the transmission as a function of applied pump frequency at $n = 1$. The full set of parameters can be found in the SI.}
    \label{fig:BCSolns}
\end{figure}

{\em Boundary effects:---} We now consider a more detailed treatment of the setup in Fig.~\ref{fig:NLExcitation} accounting for boundary conditions. The bulk equations of motion can be solved semi-analytically and a solution consistent with the boundary conditions can be found numerically by iteration; see SI for details. 
In Fig.~\ref{fig:BCSolns}, the numerically computed reflectance and transmittance coefficients of the Stokes mode are plotted as a function of index of refraction of the DA medium. We used a sample length $L = 600$ $\mu$m, so that $2QL \gg 1$, which simplifies the treatment of the nonlinearity (see SI for details).  The electric field amplitudes in the incident waves are  $E_{1a} = 5.5\times 10^{-3}$ V/nm and $E_{2a} = 10^{-6}$ V/nm; this leads to a gain factor for the Stokes mode $G_A \approx 1.32$ at $n=1$. 
The transmittance being greater than 1 is a direct consequence of  the exponential gain of the Stokes intensity governed by Eq.~\eqref{eq:I2Bulk}. Moreover, when the index of refraction $n \neq 1$, reflections at the boundaries occur, and a part of the Stokes beam can exit the sample in the direction opposite to its incidence, resulting in reflectance coefficients that can also exceed one.

Note that the Stokes and pump beam polarizations remain unchanged as they propagate through the medium if $\theta$ has no static part (A static contribution  to $\theta$ due to the equilibrium AFM order is generally present; however, even $\theta = \pi$ in topological insulators produces only a very small Kerr rotation ~\cite{ahn_theory_2022, wu_quantized_2016}. See SI for details). This simplification allows one to compute the reflection and transmission coefficients of the Stokes' mode by geometrically adding up contributions from all possible paths of the beam (with an arbitrary number of internal reflections) accounting for the phase $e^{i k_2 L}$ and gain $e^{G_A}$ factors accrued  along the segments of the path. This analytical computation matches the exact result obtained by numerical iteration well, predicting a series of Fabry-Perot type resonances where the Stokes intensity can be greatly amplified (see an example in Fig.~\ref{fig:BCSolns}, and the analytical derivation in Eqs. (S74-77) of the SI).

\begin{figure}[t]
    \includegraphics[width= 0.8\linewidth]{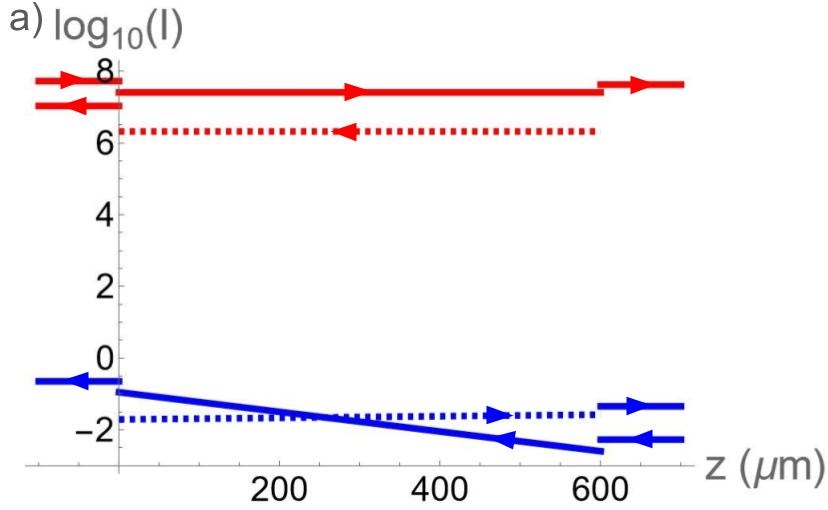}
        \includegraphics[width=0.8\linewidth]{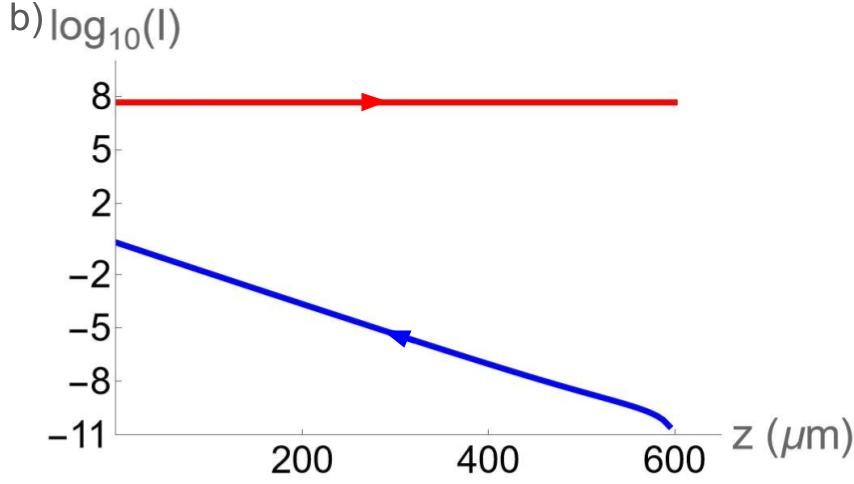}
    \caption{Log-intensities, in $\text{W}/\text{cm}^2$, of the pump (red) and Stokes (blue) beams as they traverse (and internally reflect) the DA sample. Arrows indicate propagation direction. Here we used $L = 600$ $\mu$m, $n=1.8$. Top: parametric amplification of Stokes intensity with $E_{1a} = .01$ V/nm and $E_{2a} = 10^{-7}$ V/nm and Bottom: Spontaneous generation of Stokes beam in the presence of applied pump field with $E_{1a} = 0.0125 V/nm$.
    }
    \label{fig:BCSolnsSpont}
\end{figure}


{\em Spontaneous Emission:---} 
We now explore the possibility of spontaneous generation of the Stokes mode due to the resonant amplification of quantum or thermal fluctuations in the DA medium in the presence of a single coherent source. 
Including Langevin fluctuations in Eq.~\eqref{eq:EOMtheta}, and solving for $\theta$ and $E_{2}$ in the presence of a strong and thus approximately spatially constant $E_{1}$, we find the average Stokes intensity emitted at $z=0$ (See SI) to be given by~\cite{von_foerster_quantum_1971,raymer_stimulated_1981,boyd_noise_1990}

\begin{align}
\label{eq:I2Spontaneous}
    \langle I_2(0,t=\infty)\rangle &= G_A I_{\text{Noise}} e^{G_A/2}\left[ I_0(G_A/2) - I_1(G_A/2) \right]
\end{align}
where $I_k(x)$ is the $k^{\text{th}}$ modified Bessel function of the first kind, $G_A$ is the gain factor as defined above, and $I_{\text{Noise}}$ is an effective seed Stokes intensity, 
\begin{align}
    I_{\text{Noise}} &= \frac{\epsilon_0}{4\epsilon} \coth \left( \frac{\hbar \Omega}{2 k_B T} \right) \frac{\omega_2 \gamma}{A} \approx 6 \cdot 10^{-11} \text{ W/cm$^2$}.
\end{align}
We assume $\omega_2 = 1$ THz and $\gamma = 0.04$ meV as above, sample cross-section $A \approx \lambda_2^2 = 3.55~\text{ mm}^2$ and $k_B T = 1.3$ meV. The temperature $T$ and sample cross-section $A$ enter due to the average fluctuation strength of the dynamical axion $\langle |\theta|^2\rangle \propto T/A$ (in the classical limit) as discussed in the SI.
In the large gain limit $G_A \gg 1$, Eq.~\eqref{eq:I2Spontaneous} shows that the amplitude of the Stokes mode grows exponentially with the length of the sample, much like the stimulated case. We note again that the exponential growth is limited by the condition, 
$I_2\ll I_1$, saturating when intensities become comparable. 

{\em Conclusion:---} 
SAS described in this paper is a qualitatively strong phenomenon that may be useful for spectroscopic studies of condensed matter axions, and could also have practical applications. 
For our estimates we considered Bi$_2$Se$_3$ which has well established electronic parameters, and is expected to harbor an AFM phase when doped with Fe (but has not yet been experimentally realized). The band gap and the putative N\'eel temperature are small---this limits the temperatures where the effect is present, $T < T_N$, as well as the frequencies of the electromagnetic sources, $\omega_{1,2} < |M_0|$. The above discussion also applies to thin films of $\text{MnBi}_2\text{Te}_4$, which likewise has a small band gap but a relatively large $T_N\sim 21$ K ~\cite{gao_layer_2021}. Using Eq.~\eqref{eq:J}, we find a similar gain coefficient $g_A$. We note here that bulk $\text{MnBi}_2\text{Te}_4$ possesses parity symmetry, and thus quantized $\theta$, which should make SAS vanish in bulk samples. 
%
%
SAS  should also be present in known magnetoelectrics, such as  $\text{Cr}_2 \text{O}_3$. In $\text{Cr}_2 \text{O}_3$, a significantly larger band gap (3 eV) allows application of much higher frequency pump and Stokes beams without deleterious bulk absorption, and the high Neel temperature (300 K)~\cite{iino_resistive_2019} should allow for observing SAS at room temperature. The gain coefficient benefits from higher frequency sources, though is likely suppressed by the larger band gap. An accurate estimate of $g_A$ for $\text{Cr}_2 \text{O}_3$ is left for future work.

We would like to mention that the general framework of stimulated scattering, SAS, SBS, or SRS, thanks to its exquisite selectivity and sensitivity can have many applications in condensed matter systems.  In particular, it may be useful for  detecting other long sought  excitations such as the Higgs mode \cite{
podolsky_visibility_2011, matsunaga_light-induced_2014}, or dark Josephson plasmons \cite{nicoletti_coherent_2022}. 

Finally, it is interesting to consider whether cosmic axions can be excited in a similar fashion by colliding sufficiently strong orthgonally polarized laser beams. The problem of light-by-light scattering has rich history, form the elastic Euler-Heisenberg scattering \cite{heisenberg_folgerungen_1936,karplus_scattering_1951, ueki_photon_2024}, to inelastic Beit-Wheeler electron-positron pair creation~\cite{breit_collision_1934}. At low photon frequencies, elastic scattering is extremely weak and hard to detect except in some special cases~\cite{akhmadaliev_delbruck_1998}; inelastic scattering~\cite{breit_collision_1934}, on the other hand, requires very high frequency photons~\cite{burke_positron_1997}. In this regard, SAS may be advantageous since the axion mass is expected to be in the sub eV range, and axions could be excited resonantly.
Even if such experiments do not lead to the detection of cosmic axions, they should put new bounds on the mass and the coupling to this elusive particle. 

\paragraph{Note:---}The authors of Ref.~\cite{liebman_multiphoton_2023} considered a similar set up where DAs were excited by counter-propagating electromagnetic waves. They do not include, however, the back-action of DAs on the driving fields crucial in SAS, instead studying the DA contribution to time-resolved Kerr rotations of a probe beam.

\paragraph{Data Availability:---}The code used to generate plots in Figs~\ref{fig:BCSolns} and~\ref{fig:BCSolnsSpont} can be obtained in the Harvard Dataverse with the identifier https://doi.org/10.7910/DVN/PZPUAW.

\paragraph{Acknowledgements:---}We thank V. Quito, P. Orth, S.-Y. Xu, R. McQueeney, J. Curtis, A. Bhattacharya, A. Burkov, A.  Hoffman, and D. Basov for useful discussions.
This work was supported by the Center for the Advancement of Topological
Semimetals (CATS), an Energy Frontier Research Center
funded by the U.S. Department of Energy (DOE) Office
of Science (SC), Office of Basic Energy Sciences (BES),
through the Ames National Laboratory under contract
DE-AC02-07CH11358.
The work of KA, particularly on the sample boundary effects, was
supported by the US Department of Energy, Office of Science, Basic Energy Sciences, Materials Sciences and
Engineering Division.

%



\setcounter{equation}{0}
\renewcommand{\theequation}{S\arabic{equation}}
\begin{widetext}
\section{Supplemental Information: Bulk Equations}
We start with action for electromagnetic waves and dynamical axions
\begin{align}
    \mathcal S = \int d^3x dt\frac{1}{2}(\epsilon E^2 - B^2/\mu) + \tilde\alpha\int d^3x dt (\theta_0+\theta) \bm E \cdot \bm B + J\int d^3x dt \left[ (\partial_t \theta)^2 - (v \bm\nabla \theta)^2 - m^2 \theta^2 \right]
\end{align}
where $\theta_0$ is the static part of the axion field, which in the case of magnetically doped topological insulator is the sum of the quantized part and deviations from the quantized value proportional to the AFM order. We set $\hbar = 1$ except for final expressions.
Writing $\bm E = -\partial_t \bm A$ and $\bm B = \bm\nabla \times \bm A$, $\bm A$ being the vector potential, we vary $\mathcal S$ with respect to $\bm A$ to obtain the equation of motion for $\bm A$ upon making the gauge choice $\bm \nabla \cdot \bm A = 0$:
\begin{align}
\label{eq:SI_A_EOM}
    \partial_t^2 \bm A - c'^2 \nabla^2 \bm A = \frac{\tilde\alpha}{\epsilon}\left[\dot \theta (\bm\nabla\times\bm A) - \bm\nabla\theta \times \partial_t \bm A \right] =  \frac{1}{\epsilon} \bm J_\theta
\end{align}
where
\begin{subequations}
\label{eq:Jtheta}
\begin{align}
    \bm J_\theta &= \partial_t \bm P_\theta + \bm\nabla\times \bm M_\theta \\
    \bm P_\theta &= \tilde\alpha \theta \bm B\\
    \bm M_\theta &= \tilde\alpha \theta \bm E
\end{align}
\end{subequations}
Acting upon both sides of Eq.~\eqref{eq:SI_A_EOM} with $-\partial_t$ we obtain the equation of motion for the electric field $\bm E$
\begin{align}
\label{eq:SI_E_EOM}
    \partial_t^2 \bm E - c'^2 \nabla^2 \bm E &= - \frac{1}{\epsilon} \partial_t \bm J_\theta.
\end{align}
Varying the action with respect to the  axion field $\theta$, and including damping characterized by $\gamma$, we obtain
\begin{align}
\label{eq:SI_axionEOM}
    \partial_t^2 \theta + \gamma \partial_t \theta + v^2 \nabla^2 \theta + m^2 \theta &= \frac{\tilde\alpha}{2J}\bm E \cdot \bm B
\end{align}
We will consider two counter propagating electromagnetic waves, with the total electric field being written

\begin{align}
    \bm E &= \bm E_{1}(z,t) e^{i(\omega_1 t - k_1 z)} + \bm E_{2}(z,t) e^{i(\omega_2 t + k_2 z)} + \text{ c.c.}
\end{align}
where $\bm E_{1}(z,t) = E_{1}(z,t) \hat x $ and $\bm E_{2}(z,t) = E_{2}(z,t) \hat y$ are slowly-varying envelope functions. We consider the excitation of the dynamical axion at the frequency difference $\Omega = \omega_1-\omega_2$, and we write the dynamical axion field $\theta = \theta_{12}(z,t) e^{i(\Omega t - Q z)}+$ c.c., where $Q = k_1+k_2 $
and $\theta_{12}(z,t)$ is an envelope function. We then simplify the equation of motion for the axion envelope function by working in the slowly varying envelope approximation, which neglects higher order derivatives of $\theta_{12}(z,t)$, to obtain
\begin{align}
\label{eq:SI_Axion_EOM2}
    2 i \Omega \partial_t \theta_{12}(z,t) + (m^2 - \Omega^2 - v^2 Q^2 + i \gamma 
    \Omega ) \theta_{12}(z,t) - 2i v^2 Q \partial_z \theta_{12}(z,t) &= \frac{\tilde\alpha}{2J}[E_{1}(z,t)B_{2}(z,t) + E_{2}(z,t)B_{1}(z,t)].
\end{align}
We next make the approximation that $E_{i}(z,t) \approx c' B_{i}(z,t)$, which amounts to neglecting higher order terms in $E_i(z,t)$, or equivalently higher order terms in $\tilde \alpha$ (see Eqs.~\eqref{eqs:SI_EBRelate}). We look for stationary solutions such that $\theta_{12}$ and $E_{i}$ are only functions of $z$, and using the smallness of the axion velocity in comparison to the speed of light, $v \ll c'$,
we neglect the terms proportional to $v^2$ in Eq.~\ref{eq:SI_Axion_EOM2}, so that $\theta_{12}(z)$ is given by
\begin{align}
\label{eq:SI_Theta_Soln}
    \theta_{12}(z) &= \frac{\tilde\alpha}{Jc'} \frac{E_{1}(z) E_{2}^*(z)}{m^2 - \Omega^2 + i \gamma \Omega}.
\end{align}
The polarization and magnetization arising from the dynamical axion at frequencies $\omega_1$ and $\omega_2$ are given by
\begin{subequations}
\begin{align}
    \bm P_1(z,t) &= \tilde\alpha \theta_{12}(z,t) \bm B_{2}(z,t) e^{i(\omega_1 t - k_1 z)} + \text{ c.c.}, \quad \bm M_1(z,t) =\tilde\alpha \theta_{12}(z,t) \bm E_{2}(z,t) e^{i(\omega_1 t - k_1 z)} + \text{ c.c.}\\
    \bm P_2(z,t) &= \tilde\alpha \theta_{12}^*(z,t) \bm B_{1}(z,t) e^{i(\omega_2 t + k_2 z)} + \text{ c.c.}, \quad \bm M_2(z,t) = \tilde\alpha \theta_{12}^*(z,t) \bm E_{1}(z,t) e^{i(\omega_2 t + k_2 z)} + \text{ c.c.}
\end{align}
\end{subequations}
so that 
\begin{subequations}
\begin{align}
    \bm J_1(z,t) &= 2i \omega_1 \tilde\alpha \theta_{12}(z,t) \bm B_{2}(z,t) e^{i(\omega_1 t - k_1 z)} + \text{ c.c.}\\
    \bm J_2(z,t) &= 2i \omega_2 \tilde\alpha \theta_{12}^*(z,t) \bm B_{1}(z,t) e^{i(\omega_2 t + k_2 z)} + \text{ c.c.}
\end{align}
\end{subequations}
where we have made use of $E_{i}(z,t) \approx c' B_{i}(z,t)$. Returning to Eq.~\eqref{eq:SI_E_EOM} and
using the slowly varying envelope approximation, we have the following  coupled equations for $E_{1}(z)$ and $E_{2}(z)$
\begin{subequations}
\label{eqs:SIEeom}
\begin{align}
\label{eq:SI_E1eom}
    \nabla_{z}E_{1}(z) &=- \frac{i\tilde\alpha\omega_1}{\epsilon c'^2}  \theta_{12}(z) E_{2}(z)\\
    \nabla_{z}E_{2}(z) &= \frac{i\tilde\alpha\omega_2}{\epsilon c'^2}  \theta_{12}^*(z) E_{1}(z)
\end{align}
\end{subequations}
Plugging in the expression for $\theta_{12}(z)$, Eq.~\eqref{eq:SI_Theta_Soln}, and relating the electric field amplitudes to the intensities $I_i(z) = 2 \epsilon_0 c' |E_{i}(z)|^2$, we may write the real part of Eqs.~\eqref{eqs:SIEeom} as
\begin{subequations}
\begin{align}
    \nabla_{z} I_1(z) = - g_{A} I_1(z) I_2(z)\\
    \nabla_z I_2(z) = - g_{A} \xi_{12}I_1(z) I_2(z)
\end{align}
\end{subequations}
where $\xi_{12} = \omega_2/\omega_1$ and the gain coefficient
\begin{align}
    g_{A} &= \frac{\tilde\alpha^2 \omega_1}{J \epsilon \epsilon_0 c'^4} \frac{\gamma \Omega}{(m^2- \Omega^2)^2 + (\gamma \Omega)^2}.
\end{align}
We note here that $\nabla_z I_2(z) = \xi_{12} \nabla_z I_1(z)$, which when integrated from $0$ to $z$ gives 
\begin{align}
    \frac{I_2(z)-I_2(0)}{\omega_2} &= \frac{ I_1(z)-I_1(0)}{\omega_1}
\end{align}
and is simply a statement about the conservation of photons.
We thus have
\begin{align}
    I_1(z) &= \xi_{12}^{-1}\left[ I_2(z) - I_2(0)\right] + I_1(0)
\end{align}
and the full solution for $I_2(z)$ is
\begin{align}
\label{eq:SI_I2}
    I_2(z) &= \frac{I_2(0)\left[ \xi_{12} I_1(0) - I_2(0)\right]}{\xi_{12}I_1(0) e^{g_{A}z [\xi_{12}I_1(0) - I_2(0) ]}-I_2(0)},
\end{align}
which when $\xi_{12}I_1(0) \gg I_2(0)$ reduces to Eq.~(8) in the main text.
Returning to Eqs.~\eqref{eqs:SIEeom}, we now seek the solution to the phase accumulated by the electric field envelope functions $E_{i}(z)$ due to dynamical axions. Writing $E_{i}(z) = |E_{i}(z)| e^{i\delta_i(z)}$, we have for $\delta_1(L)$
\begin{align}
    \delta_1(L) &= \delta_1(0) - \frac{g_A}{2}\frac{m^2-\Omega^2}{\gamma\Omega} \int_0^L dz\ I_2(z)
\end{align}
Evaluating the integral through use of Eq.~\eqref{eq:SI_I2}
\begin{align}
     \delta_1(L)&= \delta_1(0) - \frac{1}{2} \frac{m^2 - \Omega^2}{\gamma\Omega} \ln \left( \frac{I_2(0) e^{- g_{A} L [\xi_{12} I_1(0) -I_2(0)]} - \xi_{12}I_1(0)}{I_2(0) - \xi_{12} I_1(0)}\right)
\end{align}
which, in the limit of $\xi_{12} I_1(0) \gg I_2(0)$ is
\begin{align}
    \delta_1(L) &= \delta_1(0) - \frac{1}{2} \frac{m^2-\Omega^2}{\gamma\Omega} \frac{I_2(0)}{\xi_{12} I_1(0)}
\end{align}
Likewise, we get for $\delta_2(L)$
\begin{align}
    \delta_2(L) &= \delta_2(0) + \frac{\xi_{12} g_A}{2} \frac{m^2-\Omega^2}{\gamma\Omega} \int_0^L dz\ I_1(z) \nonumber\\
    &=\delta_2(0) + \frac{1}{2} \frac{m^2 - \Omega^2}{\gamma\Omega}\ln \left( \frac{I_2(0) - \xi_{12} I_1(0) e^{g_{A} L [\xi_{12} I_1(0)- I_2(0) ]} }{I_2(0) - \xi_{12}  I_1(0)} \right) 
\end{align}
and can be simplified in the limit $\xi_{12}I_1(0) \gg I_2(0)$
\begin{align}
    \delta_2(L) &=\delta_2(0) +  \frac{G_A}{2}\frac{m^2 - \Omega^2}{\gamma\Omega} .
\end{align}
where $G_A = g_A L \xi_{12} I_1$. As we will consider the boundary conditions for a finite system hosting dynamical axions below,
we now consider the internal reflected waves, which we write as
\begin{align}
     \bm E_{3}(z,t) e^{i(\omega_1 t + k_1 z)} + \bm E_{4}(z,t) e^{i(\omega_2 t - k_2 z)} + \text{ c.c.}
\end{align}
We assume $\hat E_{3,4} \parallel \hat E_{1,2}$, and as the internally reflected waves are propagating in the opposite direction relative to the initial internal waves, we have $\hat B_{3,4} \parallel -\hat B_{1,2}$.
We note that these internally reflected waves will also excite the dynamical axion at the frequency difference $\Omega = \omega_1 - \omega_2$, proportional to $\theta_{34}(z,t) e^{i(\Omega t + Q z)}$, where $\theta_{34}(z,t)$ is an envelope function. Returning to Eq.~\eqref{eq:SI_axionEOM}, we find for the excitation of the dynamical axion by the internal reflected waves
\begin{align}
    \theta_{34}(z) = 
    - \frac{\tilde \alpha}{Jc'}\frac{E_{3}(z) E_{4}^*(z)}{m^2 - \Omega^2 + i \gamma \Omega}
\end{align}
In writing down this solution, we have assumed we can resolve the $e^{-iQz}$ associated with $\theta_{12}(z)$ and the $e^{iQz}$ associated with $\theta_{34}(z)$, i.e. we must have $2QL \gg 1$, which we will assume going forward.
The subsequent polarization and magnetization which arise from the internally reflected waves are
\begin{subequations}
\begin{align}
    \bm P_3(z,t) &= \tilde\alpha \theta_{34}(z,t) \bm B_{4}(z,t) e^{i(\omega_1 t + k_1 z)} + \text{ c.c.} \quad \bm M_3(z,t) = \tilde\alpha \theta_{34}(z,t) \bm E_{4}(z,t) e^{i(\omega_1 t + k_1 z)} + \text{ c.c.}\\
    \bm P_4(z,t) &= \tilde\alpha \theta_{34}^*(z,t) \bm B_{3}(z,t) e^{i(\omega_2 t - k_2 z)} + \text{ c.c.} \quad \bm M_4(z,t) = \tilde\alpha \theta_{34}^*(z,t) \bm E_{3}(z,t) e^{i(\omega_2 t - k_2 z)} + \text{ c.c.}
\end{align}
\end{subequations}
which leads to currents
\begin{subequations}
\begin{align}
    \bm J_3(z,t) &= 2i\omega_1 \tilde\alpha \theta_{34}(z,t) \bm B_{4}(z,t) e^{i(\omega_1 t + k_1 z)} + \text{ c.c.}\\
    \bm J_4(z,t) &= 2i \omega_2 \tilde\alpha \theta_{34}^*(z,t)\bm B_{3}(z,t) e^{i(\omega_2 t - k_2 z)} + \text{ c.c.}
\end{align}
\end{subequations}
where again we have made the approximation to work to leading order in the fields. Returning to Eq.~\eqref{eq:SI_E_EOM} for the internally reflected waves,
the equations of motion for $E_{3,4}(z)$ are
\begin{subequations}
\label{eqs:SI_IREOM}
\begin{align}
    \nabla_z E_{3}(z) &=  -\frac{i \tilde\alpha \omega_1}{\epsilon c'^2} \theta_{34} (z) E_{4}(z)\\
    \nabla_z E_{4}(z) &= \frac{i \tilde\alpha \omega_2}{\epsilon c'^2} \theta_{34}^*(z) E_{3}(z)
\end{align}
\end{subequations}
Relating the electric field amplitudes to the intensities, the real part of the above equations can be written
\begin{subequations}
\begin{align}
    \nabla_z I_3(z) &= g_{A} I_3(z) I_4(z),\\
    \nabla_z I_4(z) &= \xi_{12} g_{A} I_3(z) I_4(z).
\end{align}
\end{subequations}
When solved, they give for the intensities of the reflected waves in the material
\begin{subequations}
\begin{align}
    I_{4}(z) &= \frac{I_4(0)\left[\xi_{12}I_3(0) - I_4(0) \right]}{\xi_{12}I_3(0) e^{-g_{A} z \left[\xi_{12} I_3(0) - I_4(0) \right]}-I_4(0)}\\
    I_3(z) &= \xi_{12}^{-1} \left[I_4(z) - I_4(0) \right] + I_3(0)
\end{align}
\end{subequations}
Note the minus sign in the exponential which assures that, even though the reflected waves are propagating in the opposite directions relative to their principle counter parts, the higher frequency mode $\omega_1$ still deposits energy into the lower frequency mode $\omega_2$. 
Returning to the imaginary part of Eqs.~\eqref{eqs:SI_IREOM} and integrating over the sample, we obtain the phases $\delta_{3,4}(L)$
\begin{subequations}
\begin{align}
    \delta_3(L) &= \delta_3(0) + \frac{1}{2} \frac{m^2 - \Omega^2}{\gamma\Omega} \ln \left( \frac{ \xi_{12} I_3(0) - I_4(0)}{\xi_{12} I_3(0) - I_4(0) e^{ g_{A} L (\xi_{12}I_3(0)- I_4(0) )}}\right)\\
    &\approx \delta_3(0) + \frac{1}{2} \frac{m^2 - \Omega^2}{\gamma\Omega} \frac{I_4(0)}{\xi_{12} I_3(0)} e^{G_A'}\\
    \delta_4(L)&= \delta_{4}(0) - \frac{1}{2}\frac{m^2 - \Omega^2}{\gamma\Omega} \ln \left( \frac{ \xi_{12} I_3(0)- I_4(0)}{ \xi_{12} I_3(0) e^{- g_{A} L ( \xi_{12} I_3(0)- I_4(0))} -I_4(0)} \right)\\
    &\approx \delta_4(0) - \frac{G_A'}{2}\frac{m^2-\Omega^2}{\gamma\Omega}
\end{align}
\end{subequations}
where we have simplified each expression in the limit $\xi_{12} I_3(0) \gg I_4(0)$, and the gain factor for the internally reflected waves is  $G_A' = g_A L \xi_{12} I_3(0)$.
\section{Parameter Estimation and Comparison to SBS}
In SI units, the axion coupling $\tilde\alpha = \alpha \epsilon_0 c/\pi$,
where $ \alpha = 1/137$ is the fine structure constant. The gain coefficient is written
\begin{align}
    g_{A} &= \frac{\alpha^2}{\pi^2} \frac{\omega_1}{J c'^2} \frac{\epsilon_0}{\epsilon}\frac{c^2}{c'^2}\frac{\gamma \hbar \Omega}{(m^2 -\hbar^2\Omega^2)^2 + (\gamma \hbar \Omega)^2}
\end{align}
here $\gamma = \hbar/\tau$ where $\tau = .1-100$ ns is the magnon lifetime~\cite{bayrakci_lifetimes_2013}. We take $\tau = .1$ ns for the estimates below.
Let us estimate $J$, and use Bi$_2$Se$_3$ doped with Fe as an example~\cite{li_dynamical_2010}. The axion stiffness $J$ can then be written
\begin{align}
    J = \tilde g^2 \tilde J = \tilde g^2 \int\frac{d^3k}{(2\pi)^3}\frac{\sum_{i=1}^4 d_i^2}{\left( \sum_{i=1}^5 d_i^2\right)^{5/2}}
\end{align}
where $d_i$ is given by
\begin{subequations}
\begin{align}
    d_1 &= \frac{A_2}{a} \sin \tilde k_x\\
    d_2 &= \frac{A_2}{a} \sin \tilde k_y\\
    d_3 &= \frac{A_1}{c} \sin \tilde k_z\\
    d_4 &= M_0 - 2\frac{B_1}{c^2}(1- \cos\tilde k_z) - 2\frac{B_2}{a^2}(1-\cos\tilde k_x) - 2\frac{B_2}{a^2}(1-\cos\tilde k_y)\\
    d_5 &= m_5
\end{align}
\end{subequations}
and $\tilde k_x = k_x a$, $\tilde k_y = k_y a$, $\tilde k_z = k_z c$.
The coupling $\tilde g$ relates the dimensionless dynamical axion to the longitudinal AFM order $m_5$, and thus $\tilde g = -M_0$~\cite{sekine_chiral_2016}. The parameters used~\cite{zhang_topological_2009}, bar $M_0$ which we assume to be small and negative so that the material is tuned to the non-topological phase, are given by
\begin{subequations}
\begin{align}
    M_0 &= -.03 \text{ eV}\\
    A_1 &= 2.2 \text{ eV$\cdot$A}\\
    A_2 &= 4.1 \text{ eV$\cdot$A}\\
    B_1 &= 10 \text{ eV$\cdot$A$^2$}\\
    B_2 &= 56.6 \text{ eV$\cdot$A$^2$}\\
    m_5 &= .001 \text{ eV}\\
    a &= 4.076 \text{ A}\\
    c &= 29.830 \text{ A}
\end{align}
\end{subequations}
The integrand is highly peaked about $(\tilde k_x, \tilde k_y, \tilde k_z) = (0,0,0)$. For ease we define
\begin{align}
    f \equiv \sum_{i=1}^5 d_i^2
\end{align}
its value at its peak is given by
\begin{align}
    f_0 =  M_0^2 + m_5^2 \approx M_0^2
\end{align}
Expanding about $(\tilde k_x, \tilde k_y, \tilde k_z) =(0,0,0)$, $f$ to second order in $\tilde k_z^2$ and $\tilde k_\parallel^2 = \tilde k_x^2  + \tilde k_y^2$ is
\begin{align}
    f &\approx f_0 + \frac{\tilde k_z^2}{c^2} \left(A_1^2 -2  M_0 B_1\right) +  \frac{\tilde k_\parallel^2}{a^2} \left( A_2^2 - 2  M_0 B_2  \right).
\end{align}
Furthermore, the integrand of $\tilde J$ is approximately
\begin{align}
    \frac{f-m_5^2}{f^{5/2}} \approx \frac{1}{f^{3/2}}
\end{align}
We thus define the relevant width of the integrand, $\Delta \tilde k$, as the dimensionless momentum for which $f$ grows by a factor of $10$, which gives
\begin{subequations}
\begin{align}
    \Delta \tilde k_z^2 &\approx \frac{10 f_0 c^2}{A_1^2 - 2  M_0 B_1},\\
    \Delta \tilde k_\parallel^2 &\approx \frac{10 f_0 a^2}{A_2^2 - 2 M_0 B_2 }.
\end{align}
\end{subequations}
The integral can then be approximated
\begin{align}
    \tilde J &\approx \int \frac{d ^3\tilde k}{(2\pi)^3 a^2 c} \frac{1}{16 f^{3/2}} \approx \frac{\Delta k_z \Delta k_\parallel^2 }{(2\pi)^3 a^2 c 16 f_0^{3/2}} \nonumber\\
    &= \frac{10^{3/2}}{16 (2\pi)^3} \frac{1}{(A_2^2 - 2 B_2 \tilde M_0)\sqrt{A_1^2 - 2 \tilde M_0 B_1}}
\end{align}
As $A_1^2 \gg 2 \tilde M_0 B_1$ and $A_2^2 \gg 2 \tilde M_0 B_2$, Eq.~\eqref{eq:SIJ} is simplified to
\begin{align}
\label{eq:SIJ}
    J &\approx \frac{10^{3/2}}{16 (2\pi)^3} \frac{M_0^2}{A_2^2 A_1} \approx 3.31 \cdot 10^{-4} \text{ eV$^{-1}$ $\cdot$ nm$^{-3}$}
\end{align}
In the case of Bi$_2$Se$_3$ doped with Fe the axion mass $m = 2$ meV, and so we choose initial frequencies
\begin{subequations}
\begin{align}
    \omega_1 &= 1483 \text{ GHz}\\
    \omega_2 &= 1000 \text{ GHz}
\end{align}
\end{subequations}
such that $\hbar\Omega = 2$ meV and we are on resonance, as well as $\hbar \omega_{1,2} \ll |M_0|$. Using these, and assuming for now $n = 1$, we have
\begin{align}
    g_A &= \frac{\tilde\alpha^2}{\pi^2} \frac{\omega_1}{J c'^2 \gamma m} = 11.6 \text{ m/GW}
\end{align}
A significant gain factor $G_A = 2\epsilon_0 c' g_{A} L \xi_{12} |E_1|^2$ is $G_A \sim 10$. For a sample of length $L = 5\cdot 10^6$ nm, so that $2 Q L \sim 10$, one obtains $G_A = 10$ with an electric field strength of
\begin{align}
    E_1 \sim .02 \text{ V/nm}
\end{align}
We note here that the gain coefficient $g_A$ can be increased by orders of magnitude. One can apply frequency $\omega_1$ which is 5 times larger and still maintain $\hbar \omega_1 < |M_0|$, and similarly we have assumed the magnon lifetime $\tau$ is small, and can be increased to $10-100$ ns.

Let us now consider stimulated Brillouin scattering (SBS)~\cite{boyd_nonlinear_2020}. Here, we consider two counter-propagating electromagnetic waves such that $\bm E_{1} \parallel \bm E_{2}$. There is a resulting electrostrictive force exerted on the sample, which drives the density $\rho$ from equilibrium, the equation of motion for $\rho$ given by
\begin{align}
\label{eq:SIrho_EOM}
     \partial_t^2 \rho + \Gamma' \partial_t \nabla_z^2 \rho - v^2 \nabla_z^2 \rho = - \frac{\gamma_e}{8\pi}\nabla_z^2 E_{tot}^2
\end{align}
Thus, the density is driven at frequency $\Omega = \omega_1 - \omega_2$, similar to the dynamical axion. 
The electric field, in turn, is driven by the change in the density as it influences the polarization:
\begin{align}
\label{eq:SI_EBrillouin_EOM}
    \partial_t^2 \bm E -c'^2 \nabla^2 \bm E = - \frac{1}{\epsilon}\partial_t^2 \bm P_B
\end{align}
where the polarization is given by
\begin{align}
\label{eq:SI_BrillouinP}
    \bm P_B &= \epsilon_0 \gamma_e \frac{ \rho}{\rho_0}\bm E.
\end{align}
Here $\gamma_e$ is a dimensionless coupling on the order of 1 and $\rho_0$ is the equilibrium density. In a similar fashion as the dynamical axion above, we
solve Eq.~\eqref{eq:SIrho_EOM} for the density $\rho(z,t)$ by writing $\rho(z,t) = \tilde\rho(z) e^{i(\Omega t - Q z)}$, where $\tilde\rho(z)$ is an envelope function, and substitute it into Eqs.~\eqref{eq:SI_BrillouinP} and~\eqref{eq:SI_EBrillouin_EOM}. 
Thus, one finds the Stokes intensity grows exponentially, similar to Eq.~\eqref{eq:SI_I2} but with a gain coefficient 
\begin{align}
    g_B = \frac{\gamma_e^2 \omega^2}{n v_{ph} c^3 \rho_0 \Gamma_B}.
\end{align}
Here it has been assumed $\omega_1 \approx \omega_2 = \omega$, $v_{ph}$ is the phonon velocity, and $\Gamma_B = \tau_{ph}^{-1}$ is the inverse phonon lifetime. Let us estimate $g_B$ for CS$_2$, which has a large gain coefficient and whose parameters are given by~\cite{boyd_nonlinear_2020}
\begin{subequations}
\begin{align}
    n &= 1.67\\
    \omega &= 6\pi\cdot 10^{14} \text{ rad/s}\\
    v &= 1.1 \cdot 10^3 \text{ m/s}\\
    \rho_0 &= 1.26 \cdot 10^3 \text{ kg/m$^3$}\\
    \gamma_e& = 2.4\\
    \Gamma_B^{-1} &= 4 \text{ ns}
\end{align}
\end{subequations}

which gives $g_B = 1.5$ m/GW. To understand why $g_A$ can be orders of magnitude larger than $g_B$, we take the ratio
\begin{align}
\label{eq:SI_ratio}
    \frac{g_A}{g_B} &= \frac{\tilde\alpha^2 n^3}{\pi^2 \gamma_e^2} \frac{\Gamma_B}{\gamma/\hbar} \frac{\omega_1}{\omega}\frac{v_{ph} c \rho_0}{\omega \hbar m J}
\end{align}
For comparison, we have separated the ratio into 4 fractions, the first 3 of which being much less than 1:
\begin{subequations}
\begin{align}
    \frac{\tilde\alpha^2 n^3}{\pi^2 \gamma_e^2} = 4.37 \cdot 10^{-6}\\
    \frac{\Gamma_B}{\gamma/\hbar} = 2.5 \cdot 10^{-2}\\
    \frac{\omega_1}{\omega} = 6.588\cdot 10^{-4}
\end{align}
\end{subequations}
And the last fraction of Eq.~\eqref{eq:SI_ratio} being $\sim 10^{12}$. To understand this, let us
re-write the last fraction in Eq.~\eqref{eq:SI_ratio} using Eq.~\eqref{eq:SIJ},
\begin{align}
    \frac{v_{ph} c \rho_0}{\omega \hbar m J} &= \frac{v^2_{ph} \rho_0}{\hbar \omega} \frac{c}{v_{ph}} \frac{1}{J m} \approx  \frac{T_D}{T_N}\frac{c}{v} \frac{8 \pi^3A_2^2 A_1 }{a^2 cM_0^2 \hbar\omega}
\end{align}
where we have written $v^2_{ph} \rho_0 \approx k_BT_D/(a^2c)$ and $m \approx k_B T_N$. These ratios are all much greater than 1, and are estimated as
\begin{subequations}
\begin{align}
    \frac{T_D}{T_N} &\approx 5 \cdot 10^2\\
    \frac{c}{v_{ph}} &\approx 10^5\\
    \frac{A_2^2A_1 }{a^2 cM_0^2 \hbar\omega} &\approx 87.9
\end{align}
\end{subequations}
so that
\begin{align}
     \frac{v c \rho_0}{\omega \hbar m J} \approx 10^{12}
\end{align}
We thus understand that the gain coefficient of SAS can be much larger than that of SBS owing to the smallness of the Neel temperature in comparison to the Debye temperature, the smallness of the phonon velocity relative to the speed of light, and the fact that $J$ is controlled by electronic parameters. 

\section{Boundary Conditions}
Let us now consider the boundary conditions of a material hosting dynamical axions in the presence of  applied counter-propagating electromagnetic waves. The electric field inside the sample is given by
\begin{align}
    \bm E &= \bm E_{1}(z,t)e^{i(\omega_1 t - k_1 z)} + \bm E_{2}(z,t) e^{i(\omega_2 t + k_2 z)} + \bm E_{3}(z,t) e^{i(\omega_1 t + k_1 z)} + \bm E_{4}(z,t)e^{i(\omega_2 t - k_2 z)} + \text{ c.c.}
\end{align}
We wish to  relate these waves with the incident waves outside the medium, $\bm E_{1,2a}$. First, let us note the Maxwell's equations in the presence of axions are~\cite{sekine_chiral_2016,sekine_axion_2021}
\begin{subequations}
\label{eqs:SI_Maxewll}
\begin{align}
    \bm\nabla \cdot \bm E &=  \frac{\rho_{charge}}{\epsilon} - \frac{1}{\epsilon}\bm \nabla \cdot \bm P_\theta \\
    \bm \nabla \times \bm E &= - \partial_t \bm B\\
    \bm \nabla\cdot\bm B &= 0\\
    \bm \nabla \times \bm B &= \frac{1}{c'^2} \partial_t \bm E +\mu \left(\bm J + \bm J_\theta\right)
\end{align}
\end{subequations}
where $\rho_{charge}$ and $\bm J$ are the charge density and current density respectively and $\bm P_\theta$ and $\bm J_\theta$ are given by Eqs.~\eqref{eq:Jtheta}.

We assume that, while our applied fields are oriented as discussed above, reflected and transmitted waves can be rotated.
The electromagnetic fields are then
\begin{subequations}
\begin{align}
    \bm E_{1} &= E_{1} (\cos \phi_1 \hat x+\sin\phi_1 \hat y), \quad &\bm B_{1} &= B_{1} (-\sin \phi_1 \hat x + \cos \phi_1 \hat y)\\
    \bm E_{2} &= E_{2}(-\sin\phi_2 \hat x + \cos\phi_2 \hat y), \quad &\bm B_{2} &= B_{2}(\cos\phi_2 \hat x+ \sin \phi_2 \hat y)\\
    \bm E_{3} &= E_{3}(\cos \phi_3 \hat x + \sin\phi_3 \hat y), \quad &\bm B_{3}& = - B_{3}(-\sin\phi_3 \hat x + \cos\phi_3 \hat y)\\
    \bm E_{4} &= E_{4}(-\sin\phi_4 \hat x + \cos\phi_4 \hat y), \quad &\bm B_{4}& = - B_{4}(\cos\phi_4 \hat x+ \sin\phi_4 \hat y)\\
    \bm E_{1a} &= E_{1a} \hat x, \quad &\bm B_{1a}& = B_{1a} \hat y\\
    \bm E_{1r} &= E_{1r}(\cos\phi_{1r} \hat x+\sin\phi_{1r} \hat y), \quad &\bm B_{1r}& = - B_{1r}(-\sin\phi_{1r} \hat x + \cos\phi_{1r}\hat y)\\
    \bm E_{1t} &= E_{1t}(\cos\phi_{1t} \hat x+\sin\phi_{1t} \hat y), \quad &\bm B_{1t}& = B_{1t}(-\sin\phi_{1t}\hat x + \cos\phi_{1t} \hat y)\\
    \bm E_{2a} &= E_{2a} \hat y, \quad &\bm B_{2a}& = B_{2a} \hat x\\
    \bm E_{2r} &= E_{2r}(-\sin\phi_{2r} \hat x + \cos\phi_{2r} \hat y), \quad &\bm B_{2r}& = - B_{2r}(\cos\phi_{2r} \hat x+\sin\phi_{2r} \hat y)\\
    \bm E_{2t} &= E_{2t}(-\sin\phi_{2t} \hat x + \cos\phi_{2t} \hat y), \quad &\bm B_{2t}& = B_{2t}(\cos\phi_{2t} \hat x + \sin\phi_{2t}\hat y)
\end{align}
\end{subequations}
As shown above, there are two contributions to the dynamical axion at the frequency difference $\Omega$, provided $2Q L \gg 1$ so that the principle and internally reflected electromagnetic waves excite the dynamical axion with amplitude $\theta_{12}$ and $\theta_{34}$ respectively,
\begin{align}
    \theta(z,t) &= \theta_{12}(z) e^{i(\Omega t - Q z)} + \theta_{34}(z)e^{i(\Omega t+Qz)} + \text{ c.c.}
\end{align}
where, in the presence of the rotated internal waves,
\begin{align}
    \theta_{12}(z) &=  \frac{\tilde\alpha}{J c'}\frac{E_{1}(z) E_{2}^*(z) }{m^2 - \Omega^2 + i \gamma\Omega } \cos(\phi_1-\phi_2)\\
    \theta_{34}(z)&= -\frac{\tilde\alpha}{Jc'}\frac{E_{3}(z)E_{4}^*(z)}{m^2 - \Omega^2 + i \gamma\Omega}\cos(\phi_3-\phi_4)
\end{align}
and we have made the approximation $E_i \approx c' B_i$.
Note that there is no contribution at the frequency difference between $E_{1}$ and $E_{4}$ because they are traveling in the same direction and thus do not excite the dynamical axion.
As we are considering the case of normal incidence, we have 
for the boundary condition for $\bm E$:
\begin{align}
    \Delta\bm E &= 0
\end{align}
Similarly for the magnetic field, we have
\begin{align}
    \Delta \bm B &= \Delta(\mu\bm M)
\end{align}
which arise from Eqs.~\eqref{eqs:SI_Maxewll}.
Here $\Delta\bm E$ is the difference between the electric fields at the interface, and similarly for $\Delta \bm B$ and $\Delta (\mu\bm M)$. The magnetization is given by
\begin{align}
    \bm M &= \frac{\mu-\mu_0}{\mu_0\mu}\bm B + \tilde\alpha (\theta_0 + \theta) \bm E
\end{align}

We will assume, for ease, that $\mu \approx \mu_0$, so the magnetization arises solely due to the static and dynamical axions and is given by
\begin{subequations}
\begin{align}
    \bm M_1(z,t) &= \tilde \alpha \theta_0 \bm E_{1}(z,t)e^{i(\omega_1 t - k_1 z)}  + \tilde \alpha \theta_0 \bm E_{3}(z,t) e^{i(\omega_1 t+ k_1 z)}+ \tilde \alpha \theta_{12}(z,t) \bm E_{2}(z,t) e^{i(\omega_1 t - k_1 z)} + \tilde\alpha\theta_{34}(z,t) \bm E_{4}(z,t) e^{i(\omega_1 t + k_1 z)} \nonumber\\
    &+\tilde\alpha\theta_{34} \bm E_{2} e^{i(\omega_1 t + (Q+k_2) z)} + \tilde\alpha\theta_{12}\bm E_{4} e^{i(\omega_1 t - (Q+k_2) z)} + \text{ c.c.}\\
    \bm M_2(z,t) &= \tilde \alpha \theta_0 \bm E_{2}(z,t) e^{i(\omega_2 t + k_2 z)} + \tilde\alpha\theta_0 \bm E_{4}(z,t)e^{i(\omega_2 t - k_2 z) } + \alpha \theta_{12}^*(z,t) \bm E_{1}(z,t) e^{i(\omega_2 t + k_2 z)} + \alpha \theta_{34}^*(z,t) \bm E_{3}(z,t) e^{i(\omega_2 t - k_2 z)} \nonumber\\
    &+ \tilde\alpha\theta_{34}^*(z,t) \bm E_{1}(z,t) e^{i(\omega_2 t - (Q+k_1)z)} + \tilde\alpha\theta_{12}^*(z,t) \bm E_{3}(z,t)e^{i(\omega_2 t + (Q+k_1) z)} + \text{ c.c.}
\end{align}
\end{subequations}
The equations stemming from $\Delta \bm E = 0$ are 
\begin{subequations}
\label{eqs:SIE_BC}
\begin{align}
    \bm E_{1a} + \bm E_{1r} &= \bm E_{1}(0) + \bm E_{3}(0)\\
    \bm E_{2t} &= \bm E_{2}(0) + \bm E_{4}(0)\\
    \bm E_{1}(L) e^{-in k_1' L} + \bm E_{3}(L) e^{in k_1' L} &= \bm E_{1t}e^{-i k_1' L}\\
    \bm E_{2}(L) e^{i nk_2' L} + \bm E_{4}(L) e^{-ink_2' L} &= \bm E_{2a} e^{ik_2' L} + \bm E_{2r}e^{-ik_2' L}
\end{align}
\end{subequations}
and the equations stemming from $\Delta \bm B = \Delta(\mu \bm M)$ are
\begin{subequations}
\label{eqs:SIB_BC}
\begin{align}
    &\bm B_{1a} + \bm B_{1r} - \bm B_{1}(0) - \bm B_{3}(0) =  - \tilde\alpha \mu_0\left[ \theta_{12}(0) + \theta_{34}(0) \right] \left[ \bm E_{2}(0) + \bm E_{4}(0) \right] - \tilde\alpha \mu_0 \theta_0 \left[ \bm E_{1}(0) + \bm E_{3}(0)  \right]\\
    &\bm B_{2t} - \bm B_{2}(0) - \bm B_{4}(0) = -\tilde\alpha \mu_0\left[ \theta_{12}^*(0) + \theta_{34}^*(0)\right]\left[ \bm E_{1}(0) + \bm E_{3}(0)\right] - \tilde\alpha \mu_0 \theta_0\left[ \bm E_{2}(0) + \bm E_{4}(0)\right]  \\
    &\bm B_{1}(L) e^{-ink_1' L} + \bm B_{3}(L) e^{ink_1' L} - \bm B_{1t}e^{-ik_1' L} =  \tilde\alpha \mu_0 \theta_{12}(L) \left[\bm E_{2}(L) e^{-ink_1' L} + \bm E_{4}(L) e^{-in(Q'+k_2')L} \right] \nonumber\\
    &+\tilde\alpha \mu_0 \theta_{34}(L) \left[\bm E_{2}(L)e^{in(Q'+k_2')L} + \bm E_{4}(L) e^{ink_1' L} \right]+ \tilde\alpha \mu_0 \theta_0 \left[\bm E_{1}(L) e^{-ink_1' L} + \bm E_{3}(L) e^{in k_1' L} \right]\\
    &\bm B_{2}(L) e^{ink_2'L} + \bm B_{4} (L) e^{-ink_2' L} - \bm B_{2a} e^{ik_2' L} -\bm B_{2r} e^{-ik_2' L} = \tilde\alpha \mu_0\theta_{12}^*(L) \left[\bm E_{1}(L) e^{in k_2' L} + \bm E_{3}(L) e^{in(Q'+k_1')L} \right]\nonumber\\
    &+ \tilde\alpha \mu_0 \theta_{34}^*(L) \left[ \bm E_{1}(L) e^{-in(Q'+k_1')L} + \bm E_{3}(L) e^{-ink_2' L}\right] + \tilde \alpha \mu_0 \theta_0 \left[ \bm E_{2}(L) e^{ink_2' L} + \bm E_{4}(L) e^{-ink_2' L}\right]
\end{align}
\end{subequations}
where $k_i' = \omega_i/c$ is the wavevector in vacuum and $Q' = k_1' + k_2'$.
Here we note that the only term that rotates the fields is proportional to $\theta_0$, as otherwise there is a consistent solution where none of the transmitted or reflected waves are rotated, i.e. $\phi_i = 0$, owing to the fact $\bm E_{1a} \parallel \bm B_{2a}$. Let us estimate the rotation due to reflection and transmission at the boundary of a semi-infinite material in the presence of $\theta_0$ and in the absence of dynamical axions. The boundary conditions in this case are 
\begin{subequations}
\begin{align}
    \bm E_{1a} + \bm E_{1r} &= \bm E_{1}(0)\\
    \bm B_{1a} + \bm B_{1r} - \bm B_{1}(0) &=  - \tilde\alpha \mu_0 \theta_0 \bm E_{1}(0)
\end{align}
\end{subequations}
We can relate the electromagnetic fields in the standard way, $E_1(0) = c' B_1(0)$ and $E_{1a,r} = c B_{1a,r}$. Solving for $\phi_1$ and $\phi_{1r}$, the angles after transmission or reflection at the boundary, we find
\begin{subequations}
\begin{align}
  \phi_1 &\approx - \frac{\alpha \theta_0/\pi}{(1+n)}  \\
  \phi_{1r} &\approx  \frac{\alpha \theta_0/\pi}{n^2 -1+ \alpha^2 \theta_0^2/\pi^2}
\end{align}
\end{subequations}
As we have thus far considered materials where the quantized part of the static axion is zero, we have $\theta_0 =- m_5/M_0$ and therefore the angle of rotation due transmission/reflection $\phi_i \ll 1$ provided $n^2 - 1 \gg \alpha^2 \theta_0^2/\pi^2$. As we are concerned with the effect of dynamical axions, we therefore neglect the term proportional to $\theta_0$ in the following and set $\phi_i = 0$. Lastly we note that even in the presence of the quantized part of the static axion, where $\theta_0 = \pi - m_5/M_0$, the rotation upon transmission or reflection is still proportional to $\alpha$, the fine structure constant, and is therefore typically negligible.

In order to simplify Eqs.~\eqref{eqs:SIB_BC}
we must relate the field amplitudes $E$ and $B$ inside the material. We do this via the Maxwell equation
\begin{align}
    \bm \nabla \times \bm E &= - \partial_t \bm B.
\end{align}
We have for the principle wave at frequency $\omega_1$
\begin{align}
    c' B_{1}(z) &= E_{1}(z)+ i \frac{\nabla_z E_{1}(z)}{k_1}
\end{align}
where we have used the fact that $\partial_t \bm B_{1}(z) = 0$. Using Eq.~\eqref{eq:SI_E1eom} we have
\begin{align}
    B_{1}(z) &= \frac{E_{1}(z)}{c'} + \tilde\alpha\mu_0 \theta_{12}(z)  E_{2}(z)
\end{align}
likewise, we get for the other waves
\begin{subequations}
\begin{align}
    B_{2}(z) &= \frac{E_{2}(z)}{c'} + \tilde\alpha \mu_0 \theta_{12}^* (z)E_{1}(z)\\
    B_{3}(z) &= \frac{E_{3}(z)}{c'} - \tilde\alpha \mu_0 \theta_{34}(z)E_{4}(z)\\
    B_{4}(z) &= \frac{E_{4}(z)}{c'} - \tilde\alpha \mu_0 \theta_{34}^*(z) E_{3}(z)
\end{align}
\end{subequations}
When plugging in the values for $\theta_{12}(z)$ and $\theta_{34}(z)$ we see
\begin{subequations}
\label{eqs:SI_EBRelate}
\begin{align}
    B_{1}(z) &= \frac{E_{1}(z)}{c'} + \frac{\tilde\alpha^2 \mu_0}{J c'}\frac{|E_{2}(z)|^2 E_{1}(z)}{m^2 - \Omega^2 + i \gamma \Omega}\\
    B_{2}(z) &= \frac{E_{2}(z)}{c'} + \frac{\tilde\alpha^2 \mu_0}{J c'}\frac{|E_{1}(z)|^2 E_{2}(z)}{m^2 - \Omega^2 - i \gamma \Omega}\\
    B_{3}(z) &= \frac{E_{3}(z)}{c'} + \frac{\tilde\alpha^2 \mu_0}{J c'}\frac{|E_{4}(z)|^2 E_{3}(z)}{m^2 - \Omega^2 + i \gamma \Omega}\\
    B_{4}(z) &= \frac{E_{4}(z)}{c'} + \frac{\tilde\alpha^2 \mu_0}{J c'}\frac{|E_{3}(z)|^2 E_{4}(z)}{m^2 - \Omega^2 - i \gamma \Omega}
\end{align}
\end{subequations}
Using Eqs.~\eqref{eqs:SI_EBRelate} our boundary conditions Eqs.~\eqref{eqs:SIE_BC} and~\eqref{eqs:SIB_BC} become
\begin{subequations}
\begin{align}
    &E_{1a} + E_{1r} = E_{1}(0) + E_{3}(0)\\
    &E_{2t} = E_{2}(0) + E_{4}(0)\\
    &E_{1}(L) e^{-i n k_1' L} + E_{3}(L) e^{in k_1' L} = E_{1t} e^{-ik_1' L}\\
    &E_{2}(L) e^{in k_2' L} + E_{4}(L)e^{-i n k_2' L} = E_{2a} e^{i k_2' L} + E_{2r}e^{-ik_2' L}\\
    &\frac{E_{1a}}{n} - \frac{E_{1r}}{n}- E_{1}(0) + E_{3}(0) = - \tilde\alpha \mu_0 c' \left[\theta_{12}(0) E_{4}(0) + \theta_{34}(0) E_{2}(0) \right]\\
    &\frac{E_{2t}}{n} - E_{2}(0) + E_{4}(0) = - \tilde\alpha \mu_0 c' \left[
    \theta_{12}^*(0) E_{3}(0) + \theta_{34}^*(0) E_{1}(0)\right]\\
    &E_{1}(L)e^{-in k_1' L} - E_{3}(L) e^{in k_1' L} - \frac{E_{1t}}{n} e^{-i k_1' L} = \tilde\alpha \mu_0 c' \left[ \theta_{34}(L)  E_{2}(L)e^{in(Q'+k_2') L} + \theta_{12}(L)  E_{4}(L) e^{-in( Q'+k_2') L}\right]\\
    &E_{2}(L) e^{in k_2' L} - E_{4}(L)e^{-in k_2' L}-\frac{E_{2a}}{n}e^{i k_2' L} + \frac{E_{2r}}{n}e^{-ik_2' L} = \tilde\alpha \mu_0 c' \left[ \theta_{34}^*(L)  E_{1}(L)e^{-in(Q'+k_1') L} + \theta_{12}^*(L)  E_{3}(L)e^{in(Q'+k_1')L}\right]
\end{align}
\end{subequations}

Note that the RHS of the latter 4 equations is only non-zero in the presence of internal reflected waves, due to the corrections to the relation between the electric and magnetic field amplitudes in the presence of dynamical axions. 
Numerical solutions for the reflectance and transmittance of the Stokes mode are plotted in Fig. 2a and 2b of the main text.

\subsection{Analytic Expression for Transmittance and Reflectance Coefficients}
Let us consider an air-sample-air system just as above, and a normally incident electromagnetic field at frequency $\omega_2$. In the absence of dynamical axions, at the first boundary, traveling from medium with index of refraction to $n_1 = 1$ to $n_2 = n$ the reflectance and transmittance coefficients are~\cite{born_principles_1999}
\begin{subequations}
\begin{align}
    r_{12} &= \frac{n_1 -n_2}{n_1+n_2}\\
    t_{12} &= \frac{2n_1}{n_1+n_2}
\end{align}
\end{subequations}

and similarly for the $23$ boundary, where $n_3 = 1$. The reflectance of the system is then given by the geometric sum of all possible internally reflected waves,
\begin{align}
    r &= r_{12} + t_{12}r_{23} t_{21} e^{2i\beta} + t_{12} r_{23} r_{21} r_{23}t_{21}e^{4i\beta} + \ldots \nonumber\\
    &= \frac{r_{12}+r_{23}e^{2i\beta}}{1+r_{12}r_{23}e^{2i\beta}}
\end{align}
here $\beta = n k_2' L$ is the phase gained upon propagation through the medium, and we have used the fact that $r_{12}^2 + t_{12}t_{21} = 1$. Similarly, we get for the transmittance
\begin{align}
    t &= t_{12}t_{23} e^{i\beta} + t_{12} r_{23} r_{21} t_{23} e^{3i\beta} + \ldots \nonumber\\
    &= \frac{t_{12}t_{23} e^{i\beta}}{1+ r_{12}r_{23} e^{2i\beta}}
\end{align}
In the presence of dynamical axions and a constant pump, which for simplicity we will assume the Stokes and pump frequencies are on resonance, $\Omega = m$, the electric field amplitude of the Stokes mode will grow exponentially in the sample. One can see how this arises by considering an imaginary contribution to the index of refraction for the principle wave, $- G_A/(2k_2' L)$. For the internally reflected wave, however, it experiences a different gain factor, $G'_A = g_A L \xi_{12} I_3(0)$, where $I_3(0)=2\epsilon_0 c' |E_{3}(0)|^2$ is the intensity of the internal reflection of the pump. Including the exponential growth correctly for the principle and reflected waves, we get for the transmission and reflection coefficients
\begin{subequations}
\begin{align}
    r &= \frac{r_{12} + r_{23} e^{2i\beta +(G_A+G'_A)/2}}{1 +r_{12}r_{23}e^{2i\beta + (G_A+G'_A)/2}}\\
    t &= \frac{t_{12}t_{21} e^{i\beta + G_A/2}}{1 +r_{12}r_{23}e^{2i\beta + (G_A+G'_A)/2}}
\end{align}
\end{subequations}
These expressions are plotted in Fig. 2 of the main text with dotted-dashed lines, with the gain factors calculated at each $n$, in excellent agreement with the numerical solutions to the boundary conditions.

\subsection{Figure Parameters}
We note here the parameters used in Figs. 2,
\begin{subequations}
\begin{align}
    J &= 3.31 \cdot 10^{-4} \text{ eV$^{-1}\cdot$nm$^{-3}$}\\
    \gamma &= \frac{\hbar}{\tau},\quad \tau = .1 \text{ ns}\\
    m &= 2 \text{ meV}\\
    \omega_1 &= 1.483 \text{ THz}\\
    \omega_2 &= 1 \text{ THz}\\
    L &= 600 \text{ $\mu$m}\\
    E_{1a} &= .0055 \text{ V/nm}\\
    E_{2a} &= 10^{-6} \text{ V/nm}
\end{align}
\end{subequations}
For the inset figure, $\omega_1$ was varied to vary $\Omega = \omega_1 -\omega_2$ at $n=1$. For Figure 3, we used the same $J$, $\gamma$, $m$, $L$, and $\omega_{1,2}$. For Figure 3a, we used
\begin{subequations}
\begin{align}
    E_{1a} &= .01 \text{ V/nm}\\
    E_{2a} &= 10^{-7} \text{ V/nm}\\
    n &= 1.8
\end{align}
\end{subequations}
for Fig. 3b we use the same index of refraction $n$ but $E_{1a} = .0125$ V/nm.

\section{Spontaneous Axion Scattering}
Let us consider the spontaneous generation of the Stokes field in the presence of a pump field and a fluctuating $\theta$ (AFM order).
We include in our equation of motion for $\theta$ a Langevin noise term $F(z,t) = i\Omega f(z,t) e^{i(\Omega t - Q z)}$~\cite{von_foerster_quantum_1971,raymer_stimulated_1981,boyd_noise_1990}, so that the equation of motion is
\begin{align}
    \partial_t^2 \theta + \gamma \partial_t \theta + v^2 \nabla^2 \theta + m^2 \theta = \frac{\tilde \alpha}{2J} \bm E \cdot \bm B + F
\end{align}
which, when plugging in our expression for the counter-propagating electromagnetic waves $E_{1}$ and $E_{2}$ and working in the slowly varying envelope approximation, gives
\begin{align}
    2i \Omega \partial_t \theta_{12}(z,t) + (m^2 - \Omega^2 - v^2 Q^2 + i \gamma\Omega) \theta_{12}(z,t) - 2 i v^2 Q \nabla_z \theta_{12}(z,t) = \frac{\tilde\alpha}{Jc'}E_{1}(z,t)E_{2}^*(z,t) + i \Omega f(z,t)
\end{align}
Let us further simplify by neglecting terms proprotional to $v^2$ and considering the response on resonance, i.e. $\Omega = m$,
\begin{align}
    \partial_t \theta_{12}(z,t) + \frac{\gamma}{2} \theta_{12}(z,t) &= \frac{\tilde\alpha}{2J c' \Omega}E_{1}(z,t)E_{2}^*(z,t) + \frac{ f(z,t)}{2}.
\end{align}
In the absence of $E_{1,2}$ this is easily solved by writing $\tilde \theta = \theta e^{\gamma t/2}$ and integrating over time. The resulting solution is
\begin{align}
    \theta_{12}(z,t) &= \int_{-\infty}^t dt' \frac{f(z,t')}{2} e^{\gamma(t'-t)/2}
\end{align}
where we have used the fact that $\theta_{12}(z,-\infty) = 0$. Let us partition our sample into rectangles of length $\Delta z$ and cross sectional area $A$, with the value of $\theta_{12}(z,t)$ and $f(z,t)$ in the $i-$th rectangle denoted $\theta_i(t)$ and $f_i(t)$  respectively. The correlator $\langle \theta_i(t) \theta^*_j(t) \rangle$ is then 
\begin{align}
    \langle \theta_i(t) \theta_j^*(t) \rangle =\frac{1}{4} \int dt' dt'' \langle f_i(t') f_j^*(t'')\rangle e^{\gamma (t'+t''-2t)/2} 
\end{align}
if we assume the correlations of the Langevin noise take the form
\begin{align}
    \langle f_i(t) f_j^*(t') \rangle =\tilde Q  \delta_{ij}\delta(t-t')
\end{align}
we then have for the correlator of $\theta$
\begin{align}
\label{eq:thetacorrelator}
    \langle \theta_i(t) \theta_j^*(t) \rangle = \frac{\tilde Q}{4\gamma} \delta_{ij}
\end{align}
The fluctuations of $\theta$ can be estimated as follows: the dynamical axion field inside each box, $\theta_i$, has average energy density
\begin{align}
    \langle u_i \rangle &= J \langle |\dot \theta_i|^2 \rangle  + Jm^2 \langle |\theta_i|^2 \rangle
\end{align}
The total average energy of each box is then $\mathcal E_i = \langle u_i \rangle A \Delta z$. By the equipartition theorem, each term contributes an energy $k_B T/2$, and we therefore estimate
\begin{align}
    \langle |\theta_i|^2 \rangle &= \frac{k_B T}{2J m^2 A \Delta z}
\end{align}
which gives us for $\tilde Q$
\begin{align}
    \tilde Q = \frac{2k_B T \gamma}{ J m^2 A \Delta z}
\end{align}
writing the correlator for the Langevin noise in the continuum limit, we have
\begin{align}
\label{eq:fcorrelator}
    \langle f(z,t) f^*(z',t') \rangle = Q \delta(z-z') \delta(t-t')
\end{align}
where we have written $Q = \tilde Q \Delta z$ and 
\begin{align}
     Q = \frac{2k_B T \gamma}{ J m^2 A}
\end{align}
For the consideration of quantum fluctuations, one can replace $k_B T$ with $\frac{\hbar \Omega}{2}(2\bar n +1)$, where $\bar n = (e^{\hbar\Omega/(k_BT)}-1)^{-1}$ is the occupation number of magnons, and $Q$ is therefore given by

\begin{align}
    Q &= \frac{\gamma \hbar \Omega}{J m^2 A}\coth\left( \frac{\hbar \Omega}{2 k_B T}\right).
\end{align}
Note that $Q$ reduces to the expression for thermal fluctuations in the limit $k_B T \gg \hbar \Omega$.
We now solve the equations for $E_{2}$ and $\theta$, considering $E_{1}$ to be constant in space and time. The relevant equations are
\begin{subequations}
\begin{align}
    (\nabla_z - \frac{1}{c'}\partial_t) E_{2}(z,t) = \frac{\tilde\alpha\omega_2}{i \epsilon c'^2}\theta^*_{12}(z,t) E_{1}\\
    \partial_t \theta^*_{12}(z,t) + \frac{\gamma}{2}\theta^*_{12}(z,t) = \frac{i\tilde\alpha E_{1}^* }{2 J c' \Omega} E_{2}(z,t) + \frac{f^*(z,t)}{2}
\end{align}
\end{subequations}
for simplicity, we write 
\begin{subequations}
\begin{align}
    \kappa_1 &= \frac{\tilde\alpha \omega_2}{\epsilon c'^2} E_{1}\\
    \kappa_2 &= \frac{\tilde\alpha}{2J c' \Omega}E_{1}^*
\end{align}
\end{subequations}
and change variables to $(z, \tau)$ where $\tau = t + z/c'$
which allows us to write our equations
\begin{subequations}
\begin{align}
    \partial_x E_{2}(z,\tau) &= -i \kappa_1 \theta^*_{12}(z,\tau)\\
    \partial_\tau \theta^*(z,\tau) + \frac{\gamma}{2}\theta^*_{12}(z,\tau) &= i \kappa_2 E_{2}(z,\tau) + \frac{f^*(z,\tau)}{2}
\end{align}
\end{subequations}
We now Laplace transform both equations in coordinate $z$, $\mathcal L\left\{ g(z) \right\} = \int_0^\infty e^{-sz} g(z) dz$ , using the notation
\begin{subequations}
\begin{align}
    \mathcal L \left\{ E_{2}(z,\tau) \right\} &= \tilde E_{2}(s,\tau)\\
    \mathcal L \left\{ \theta^*_{12}(z,\tau) \right\} &= \tilde \theta_{12}^*(s,\tau)\\
    \mathcal L \left\{ f^*(z,\tau) \right\} &=\tilde f^*(s,\tau)
\end{align}
\end{subequations}
to obtain
\begin{subequations}
\begin{align}
    \tilde E_{2}(s,\tau) &= s^{-1}\left[E_{2}(0,\tau) -i \kappa_1 \tilde\theta_{12}^*(s,\tau)\right]\\
    \partial_\tau \theta^*_{12}(s,\tau) + \frac{\gamma}{2} \tilde\theta^*_{12}(s,\tau) &= i\kappa_2 \tilde E_{2}(s,\tau) + \frac{\tilde f^*(s,\tau)}{2}.
\end{align}
\end{subequations}
Plugging the former into the latter we can solve for $\theta^*_{12}(s,\tau)$
\begin{align}
    \tilde\theta_{12}(s,\tau) &= \tilde\theta_{12}(s,0) e^{-\gamma\tau/2}e^{\kappa_1\kappa_2 \tau/s}+ \int_0^\tau d\tau' e^{-\gamma(\tau-\tau')/2}e^{\kappa_1\kappa_2(\tau-\tau')/s}\left( \frac{i\kappa_2 E_{2}(0,\tau')}{s}+ \frac{\tilde f^*(s,\tau')}{2} \right)
\end{align}
which we then plug into the equation for $\tilde E_{2}(s,\tau)$ to obtain
\begin{align}
\label{eq:LaplaceE2}
    \tilde E_{2}(s,\tau) &= \frac{E_{2}(0,\tau)}{s} - i \kappa_1 \tilde\theta^*_{12}(s,0)e^{-\gamma\tau/2}\frac{e^{\kappa_1\kappa_2 \tau/s}}{s} -\frac{i\kappa_1}{s} \int_0^\tau d\tau' e^{-\gamma(\tau-\tau')/2}e^{\kappa_1\kappa_2(\tau-\tau')/s}\left( \frac{i\kappa_2 E_{2}(0,\tau')}{s}+ \frac{\tilde f^*(s,\tau')}{2} \right).
\end{align}
Taking the inverse Laplace transform of Eq.~\eqref{eq:LaplaceE2}, we make use of the following two identities:
\begin{subequations}
\label{eq:identities}
\begin{align}
    \mathcal L\left\{(z/a)^n I_n(\sqrt{4 a z}) \right\} &= s^{-(n+1)}e^{a/s},\\
    \mathcal L^{-1}\left\{ F(s)G(s) \right\} &= \int_0^x f(x-z)g(z)dz,
\end{align}
\end{subequations}
where $I_n(x)$ is the $n-$th order modified Bessel function and $f$ and $g$ are the inverse Laplace transforms of $F$ and $G$. Using identities Eqs.~\eqref{eq:identities}, we find for $E_2(z,\tau)$
\begin{align}
    E_{2}(z,\tau) &= E_{2}(0,\tau) - i \kappa_1 e^{-\gamma \tau/2} \int_0^z dz' \theta^*_{12}(z',0) I_0(\sqrt{4\kappa_1\kappa_2 \tau(z-z')}) \nonumber\\
    &+ \sqrt{\kappa_1\kappa_2 z}\int_0^\tau d\tau' \frac{e^{-\gamma(\tau-\tau')/2}}{\sqrt{\tau-\tau'}}E_{2}(0,\tau') \nonumber\\
    &-i\frac{\kappa_1}{2} \int_0^\tau d\tau'\int_0^z dz' e^{-\gamma(\tau-\tau')/2} f^*(z',\tau')I_0(\sqrt{4\kappa_1\kappa_2(\tau-\tau')(z-z')})
\end{align}
Let us set $z = L$ and use the fact that $|E_{2}(L,\tau)| \ll |E_{2}(0,\tau)|$, as we know $E_2$ will grow from $0$ to $L$. We can then multiply the above equation by its complex conjugate and do a statistical average to obtain
\begin{align}
\label{eq:SpontE2v1}
    &\langle|E_{2}(0,\tau)|^2 \rangle = |\kappa_1|^2 e^{-\gamma\tau} \int_0^L dz' dz''  \langle \theta_{12}^*(z',0) \theta_{12}(z'',0)\rangle I_0(\sqrt{4\kappa_1\kappa_2 \tau(L-z')})I_0(\sqrt{4\kappa_1\kappa_2 \tau(L-z'')}) \nonumber\\
    &+ \frac{|\kappa_1|^2 }{4}\int_0^\tau d\tau'd\tau'' \int_0^L dz' dz'' e^{-\gamma(2\tau - \tau'-\tau'')/2}\langle f^*(z',\tau') f(z'',\tau'') \rangle I_0(\sqrt{4\kappa_1\kappa_2 (\tau-\tau')(L-z')})I_0(\sqrt{4\kappa_1\kappa_2 (\tau-\tau'')(L-z'')})
\end{align}
where we have assumed that $E_{2}(0,\tau)$ is not correlated with $f(z,\tau)$, $\theta_{12}(z,\tau)$, or itself at different times, and that $\langle \theta_{12}(z,\tau) f^*(z',\tau') \rangle = 0$.
To simplify Eq.~\eqref{eq:SpontE2v1} we use Eq.~\eqref{eq:fcorrelator} and the continuum version of Eq.~\eqref{eq:thetacorrelator},
\begin{align}
    \langle\theta(z,0) \theta^*(z',0) &= \frac{Q}{4\gamma} \delta(z-z'),
\end{align}
giving
\begin{align}
\label{eq:SI_SpontStokesE2}
    \langle |E_{2}(0,\tau)|^2\rangle &= \frac{|\kappa_1|^2 Q L}{4\gamma} e^{-\gamma\tau} \left[I_0^2(\sqrt{4\kappa_1\kappa_2 \tau L}) - I_1^2(\sqrt{4\kappa_1\kappa_2 \tau L})\right] \nonumber\\
    &+\frac{|\kappa_1|^2}{4} Q L \int_0^\tau d\tau' e^{-\gamma(\tau-\tau')}\left[I_0^2(\sqrt{4\kappa_1\kappa_2 (\tau -\tau')L}) - I_1^2(\sqrt{4\kappa_1\kappa_2 (\tau-\tau') L})\right].
\end{align}
At long times $\gamma\tau \gg 1$ the first term can be neglected. To see this, we use that at large arguments the modified Bessel functions can be written
\begin{subequations}
\begin{align}
    I_0(\sqrt{a \tau}) &\approx \frac{e^{\sqrt{a\tau}}}{\sqrt{2\pi}}\left[(a\tau)^{-1/4} + \frac{(a\tau)^{-3/4}}{8} + \ldots\right]\\
    I_1(\sqrt{a \tau}) &\approx \frac{e^{\sqrt{a\tau}}}{\sqrt{2\pi}}\left[(a\tau)^{-1/4} -\frac{3(a\tau)^{-3/4}}{8} + \ldots\right]
\end{align}
\end{subequations}
Thus, when paired with the exponential $e^{-\gamma\tau}$ the first term of Eq.~\eqref{eq:SI_SpontStokesE2} will always go to zero in the long time limit. To see that the third term is finite, we change variables to $y = \gamma( \tau-\tau')$,
\begin{align}
     \langle |E_{2}(0,\tau)|^2\rangle &\approx -\frac{|\kappa_1|^2 Q L}{4\gamma} \int_{\gamma\tau}^0 dy\ e^{-y}\left[ I_0^2\left(\sqrt{\frac{4\kappa_1\kappa_2L}{\gamma} y} \right) - I_1^2\left(\sqrt{\frac{4\kappa_1\kappa_2L}{\gamma} y}\right)\right]
\end{align}
in the limit that $\gamma \tau \gg 1$, or $\tau \rightarrow \infty$, we have 
\begin{align}
    \langle |E_{2}(0,\infty)|^2\rangle &= \frac{|\kappa_1|^2 Q L}{4\gamma} \int_{0}^\infty e^{-y}\left[ I_0^2\left(\sqrt{\frac{4\kappa_1\kappa_2L}{4\gamma} y}
    \right)- I_1^2\left(\sqrt{\frac{4\kappa_1\kappa_2L}{\gamma} y}\right)\right] \nonumber\\
    &= \frac{|\kappa_1|^2 Q L}{4\gamma}e^{\frac{2\kappa_1\kappa_2 L}{\gamma}} \left[ I_0\left(\frac{2\kappa_1\kappa_2 L}{\gamma}\right) - I_1\left(\frac{2\kappa_1\kappa_2 L}{\gamma}\right)\right]
\end{align}
we can rewrite
\begin{align}
    \frac{2\kappa_1\kappa_2 L}{\gamma} &= \frac{\alpha^2}{\pi^2}\frac{\omega_2\epsilon_0}{J c'}\frac{\epsilon_0}{\epsilon}\frac{c^2}{c'^2}\frac{|E_{1}|^2 L}{\gamma\Omega} = 2\epsilon_0 c' g_{A} L \xi_{12} |E_{1}|^2/2 = G_A/2
\end{align}
where we have used the expression for $ g_{A}$ on resonance. The average Stokes intensity at $\gamma \tau \gg 1$ is then
\begin{align}
\label{eq:SI_I2Spont}
    \langle I_2(0,\infty)\rangle &= G_A  I_{\text{Noise}} e^{G_A/2}\left[ I_0(G_A/2) - I_1(G_A/2) \right]
\end{align}
with 
\begin{align}
    I_{\text{Noise}} &=  \frac{\epsilon_0}{4\epsilon} \coth\left(\frac{\hbar\Omega}{2k_B T} \right)\frac{\omega_2 \gamma}{A} \approx 6 \cdot 10^{-11} \text{ W/cm$^2$}
\end{align}
where we have used $\omega_2 = 1000$ GHz, $\gamma = .04$ meV, $n=1$ as well as using a sample size $A \sim \lambda_2^2 = 3.55$ mm$^2$ and $k_B T \sim 1.3$ meV. We note however that Eq.~\eqref{eq:SI_I2Spont} is valid under the assumption that $E_{1}$ is constant, and thus Eq.~\eqref{eq:SI_I2Spont} is only valid when $\langle I_2(0,\infty)\rangle \ll I_1$.

\end{widetext}
\end{document}